\documentclass[twocolumn,amsmath,twocolumn,amssymb,floatfix,showpacs,superscriptaddress]{revtex4-2}

\usepackage{graphicx}
\usepackage{bm}
\usepackage{enumerate}
\usepackage{color}
\usepackage{float}
\usepackage{bbold}
\usepackage{nccmath}
\usepackage{appendix}
\usepackage{amsmath}
\usepackage{changepage}
\usepackage{caption}
\usepackage{subcaption}
\usepackage{epstopdf}
\usepackage{tikz}
\usepackage{stfloats}

\usepackage[percent]{overpic} 
\usepackage{braket}
\usepackage[colorlinks=true,linkcolor=blue,citecolor=blue,urlcolor=blue]{hyperref}

\begin{document} 
\title{Magnetotransport properties of an unconventional Rashba spin-orbit coupled two-dimensional electronic system} 

\author{Aryan Pandita}
\author{SK Firoz Islam}
\email{s\_islam2@jmi.ac.in}
\affiliation{Jamia Millia Islamia, New Delhi-110025, INDIA}

\begin{abstract}
We study the magnetotransport properties of a two-dimensional electronic system with unconventional Rashba spin-orbit coupling in which the system is described by a pair of chiral spin textures in each spin branch, and the chirality is opposite in the two spin branches. We obtain the Landau levels analytically and find that intra-spin and/or inter-spin Landau level crossing occurs. We compute the longitudinal conductivity and quantum Hall conductivity using the Kubo formalism based on linear response theory. We find that the usual Shubnikov-de Haas oscillation in longitudinal conductivity can be made purely spin polarized by adjusting the Fermi level suitably. We observe a beating pattern in the Shubnikov-de Haas oscillation in the intra-spin branches, which arises due to the superposition of Shubnikov-de Haas oscillations corresponding to two bands in each spin branch. This is contrary to the conventional Rashba system, where such beating is due to the superposition of Shubnikov-de Haas oscillations corresponding to the two spin-branches. On the other hand, we note that quantum-Hall conductivity exhibits usual quantization in units of $e^2/h$ corresponding to each spin-dependent Landau level. However, the Landau level crossing gives rise to the double jump in the Hall conductivity if the Fermi level is placed precisely at the crossing point.

\end{abstract}

\maketitle

\section{Introduction}
The spin-orbit interaction (SOI) is a relativistic effect in a fermionic system that is the key ingredient in the field of spintronics, including several exciting quantum phenomena like spin-Hall effects \cite{Sinova2015}, Edelstein effect \cite{Edelstein1990}, Persistent spin helix \cite{Koralek2009} in semiconductor heterostructures, and $Z_2$ topological insulation \cite{Kane2005,Hasan2010} in Dirac materials, etc. Apart from the semiconductor heterostructure \cite{Sinova2015,Das1989}, the SOI is also found in metal surfaces and interfaces \cite{Ast2007, Sanchez2013,Hibino2021, Isasa2016,Nakayama2016, Miron2011,Miron2010} as well as low-dimensional materials \cite{Mourik2012,Zhang2017}. There are generally two types of spin-orbit interaction in semiconductor heterostructures: Rashba spin-orbit interaction (RSOI) and Dresselhaus spin-orbit interaction (DSOI) \cite{Winkler2003}. The RSOI arises mainly due to the lack of inversion symmetry in the quantum well, whereas the DSOI is due to the bulk inversion asymmetry \cite{Winkler2003}. One of the unique aspects of the RSOI or DSOI is that it can remove the spin degeneracy even without a magnetic field. The spin texture of each spin branch is opposite in chirality at a specified Fermi level. There are a number of compounds like $\rm BiTeI$, $\rm GaInAs/InP$, have been found to exhibit RSOI \cite{Ishizaka2011,Nitta1997,Engels1997,Grundler2000} and DSOI \cite{Meier2007,Koralek2009,Ganichev2004,Dettwiler2017}. In the early 90s, RSOI and DSOI were confirmed experimentally \cite{Das1989}. In fact, the RSOI was found to be significantly enhanced by the application of gate voltage, whereas DSOI is an intrinsic bulk property and cannot be modulated by external gate voltage \cite{Nitta1997,Engels1997,Grundler2000}. 

In recent times, a new type of Rashba spin-orbit interaction has been under consideration from the theoretical perspective, also known as unconventional RSOI (URSOI) \cite{Wang2024,Huang2024,Bhattacharya2025}. In this case, each spin branch exhibits two bands, and spin degeneracy remains lifted even at $\Gamma$ point. Contrary to the conventional Rashba system, here the spin textures of the two bands in each spin branch have the same chirality but are opposite to those of the other spin branch. Few systems have been reported to exhibit the unconventional RSOI, namely ${\rm Bi/Cu}$ (111) \cite{Mirhosseini2009} , monolayer ${\rm OsBi_2}$ \cite{Song2021} and ${\rm BiAg_2}$ \cite{Nechaev2019}. The underlying physical origin for such an unconventional nature of RSOI is still not fully understood, but it is believed to be the result of the mixing of bands and orbitals. Unlike the conventional Rashba coupling term that was originally proposed by E. I. Rashba \cite{Bychkov1984}, the complete analytical form of the unconventional Rashba coupling term is still lacking. However, recently Huang et al.  \cite{Huang2024} have attempted to explain such a system by using a simple {${\bf k}\cdot {\bf p}$} model that has allowed to investigate the microscopic theory of superconductivity \cite{Wang2024}, second-order anomalous Hall transport \cite {Bhattacharya2025}, etc.

In this work, we aim to investigate the magnetotransport properties of such a system in the quantum regime, particularly the integer quantum-Hall effect (IQHE). The IQHE is the quantization of Hall conductivity at very low temperature of a two-dimensional electronic system that was discovered by Klitzing et. al. in the early 80s \cite {Klitzing1980},  this phenomena is also considered to be the basic building block of  newly emerged field-band topology.  The IQHE has received considerable attention following the experimental observation of it at the room-temperature in graphene \cite{Novoselov2007,Zhang2005,Gusynin2005}. Subsequently, a series of investigations were carried out in different kinds of newly emerged Dirac type materials, as gapped graphene \cite{krstajic2012} bilayer graphene \cite{Novoselov2006}, silicene \cite{Tahir2013,Shakouri2014}, molybdenum disulfide (${\rm MoS_2}$) \cite{Tahir2016}, tilted Dirac material-like 8-pmmn borophene \cite{Islam2018}, semi-Dirac material \cite{Sinha2020}, $\alpha-T_3$ lattice \cite{Biswas2016} etc. The IQHE was also theoretically investigated in a semiconductor heterostructure with conventional RSOI \cite{Wang2003,Firoz_Rashba}, and in the presence of both the RSOI and DSOI \cite{Wang2005, Zhang2006}.

In this work, we study the IQHE for an electronic system with URSOI. We analytically obtain the Landau levels and show intra-spin and/or inter-spin Landau level crossing occurs. We also evaluate longitudinal conductivity and Hall conductivity using the Kubo formalism. The Landau level crossing is found to cause a double jump in the steps of Hall conductivity when the Fermi level is precisely placed at that crossing. As two spin branches are well separated in energy space, we find that by suitably choosing the Fermi level, a pure spin-polarized longitudinal conductivity or quantum Hall conductivity can be observed. 

This paper is organized as follows. In Sec.\ref{Model hamiltonian} we present the band structure, derivation of Landau levels, and the corresponding Density of states (DOS). In Sec.\ref{longandhall}, we evaluate the longitudinal and hall conductivities, followed by the summary and conclusion in Sec.\ref{Conclusion}
\vspace{0cm}
\section{Model Hamiltonian}\label{Model hamiltonian}
\begin{figure}[H]
    \centering  
    \hspace*{-.1cm}
    \begin{minipage}{0.5\textwidth}
        \includegraphics[width=8.2cm,height=6cm]{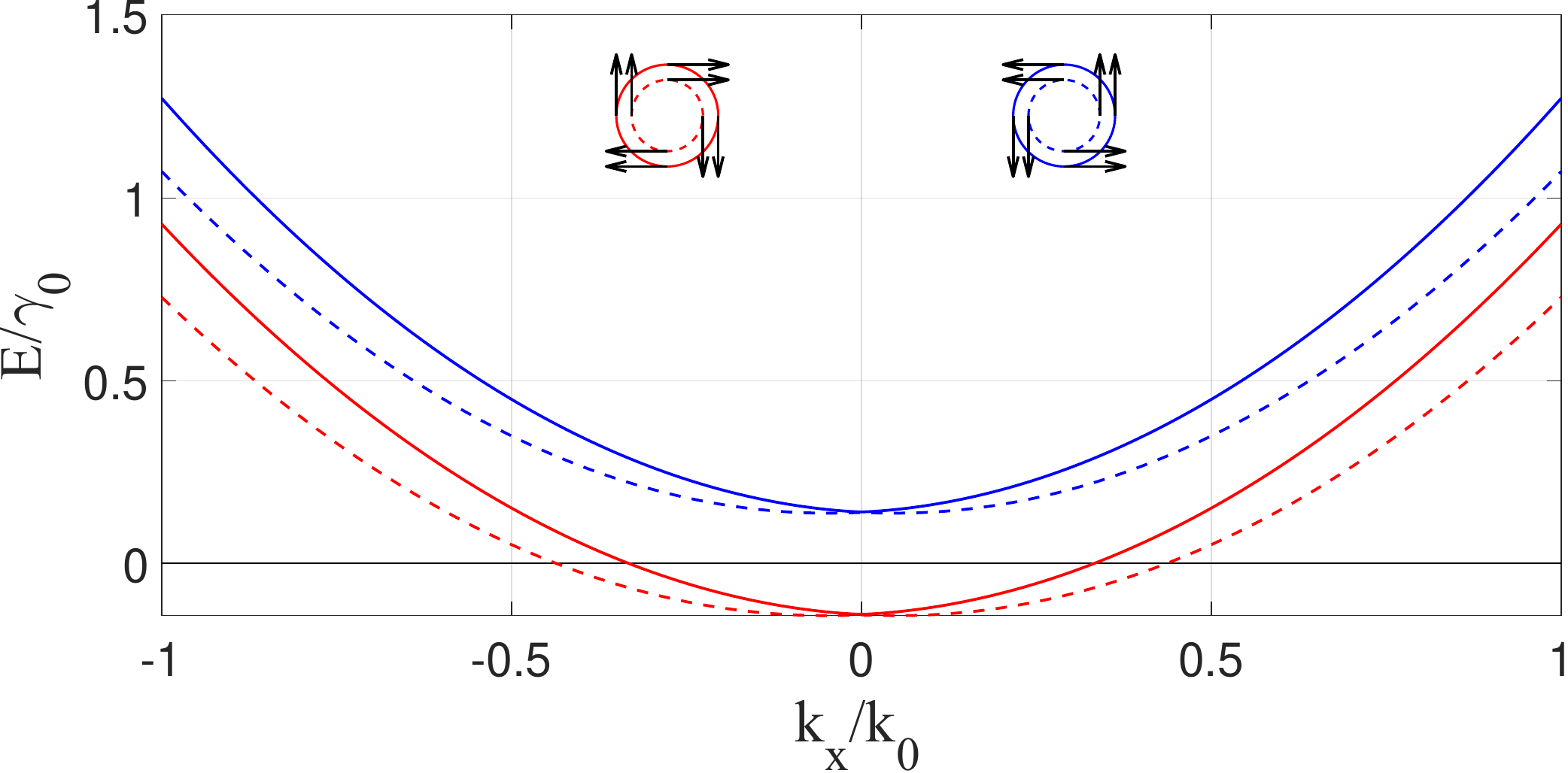}
        \caption{Energy dispersion is plotted for the parameter \({\alpha}=0.1\) and \({\eta}=0.1\) that are normalized by a typical energy scale $\gamma_0=\hbar^2k_0^2/2m^*$ with $k_0$ being a wave vector corresponding to standard $2$D electron density . Blue and red lines denote two spin branches. Each spin branch again consists of two bands, denoted by dashed and solid lines. .}
       \label{fig:bandspectrum}
    \end{minipage}
\end{figure}

Let's consider the $2$D electronic system with URSOI, that lies in the $xy$ plane, described by the low-energy effective Hamiltonian as \cite{Wang2024,Huang2024,Bhattacharya2025}
\begin{equation}
H= \frac{\bf p^2}{2m^*}\,\tau_{0}\sigma_{0} 
- \frac{\alpha}{\hbar} \left( \tau_{0} + \tau_{1} \right)\, (\boldsymbol{\sigma} \times \mathbf{p})_{z} 
+ \eta\, \tau_{2}\sigma_{3}.
\end{equation}
where ${\bf \sigma} \equiv \{\sigma_x, \sigma_y,\sigma_z\}$, ${\bf \tau}=\{\tau_1,\tau_2,\tau_3\}$ are Pauli matrices in real and orbital spin space, respectively. Whereas, $\tau_{0}$ and $\sigma_{0}$ are the $2 \times 2$ identity matrices, while \(\alpha\) represents the strength of the Rashba spin–orbit interaction (RSOI), \(\eta\) corresponds to the onsite spin–orbit coupling and $m^*$ is the effective mass of electron. The above Hamiltonian can be diagonalized to obtain the energy spectrum as  
\begin{equation}\label{dispersion}
E_{k,s_1,s_2} = \frac{\hbar^2k^2}{2m^*} + s_1 \sqrt{\eta^2 + k^2 \alpha^2} + s_2 \alpha |k| ,
\end{equation}
where $s_1 = +(-) $ corresponding to spin polarization $\uparrow(\downarrow)$, and $s_2=\pm$ denotes the two bands corresponding to two orbitals, respectively. The energy dispersion is plotted in Fig.~(\ref{fig:bandspectrum}) showing two bands in each spin branch.  

The average spin polarization can be obtained as $\langle {\bf S}\rangle=\langle \Psi_{\zeta}|{\bf S}|\Psi_{\zeta}\rangle$ where $\Psi_\zeta=\exp(i {\bf k\cdot r})[c_{k,1\uparrow}~c_{k,1\downarrow}~c_{k,2\uparrow}~c_{k,2\downarrow}]^T/\sqrt{\Omega}$ as $S_{s_1, s_2} = s_1 \alpha (-k_y e_x + k_x e_y) / \sqrt{\eta^2 + \alpha^2 k^2}$ \cite{Huang2024}. In the Fig.~(\ref{fig:bandspectrum}), the blue and red lines denote two spin branches with opposite spin polarization. whereas the dashed and solid lines represent two different bands. 
 Additionally, spin splitting remains even at the $\Gamma$ point, which is in complete contrast to conventional Rashba spin-orbit coupled electronic systems \cite{Winkler2003}.
\subsection{Inclusion of magnetic field}
Now the magnetic field (${\bf B}=B\hat{z}$) is normally applied to the electronic system, which is included via the Landau-Peierls substitution ${\bf p}\rightarrow {\bf p+eA}$ into the Hamiltonian as
\begin{equation}
H= \frac{(\mathbf{p} + e \mathbf{A})^2}{2m^*}\,\tau_{0}\sigma_{0} 
- \frac{\alpha}{\hbar} \left( \tau_{0} + \tau_{1} \right)\, \big[\boldsymbol{\sigma} \times (\mathbf{p} + e \mathbf{A})\big]_{z} 
+ \eta\, \tau_{2}\sigma_{3}.
\end{equation}
To proceed further we use  the Landau gauge ${\bf A}=(0,xB,0)$ which yields
\begin{align}
H &= \frac{1}{2m^*} \Bigl[p_x^2 + (p_y + eBx)^2\Bigr] \tau_{0}\sigma_{0} \nonumber \\[4pt]
&\quad - \frac{\alpha}{\hbar} (\tau_{0} + \tau_{1})
   \Bigl[\sigma_x (p_y + eBx) - \sigma_y p_x\Bigr]
   + \eta\,\tau_{2}\sigma_z.
\label{H}
\end{align}

\begin{figure}[h!]
    \hspace{0.0cm}
    \begin{minipage}{0.47\textwidth}
        \includegraphics[width=\linewidth]{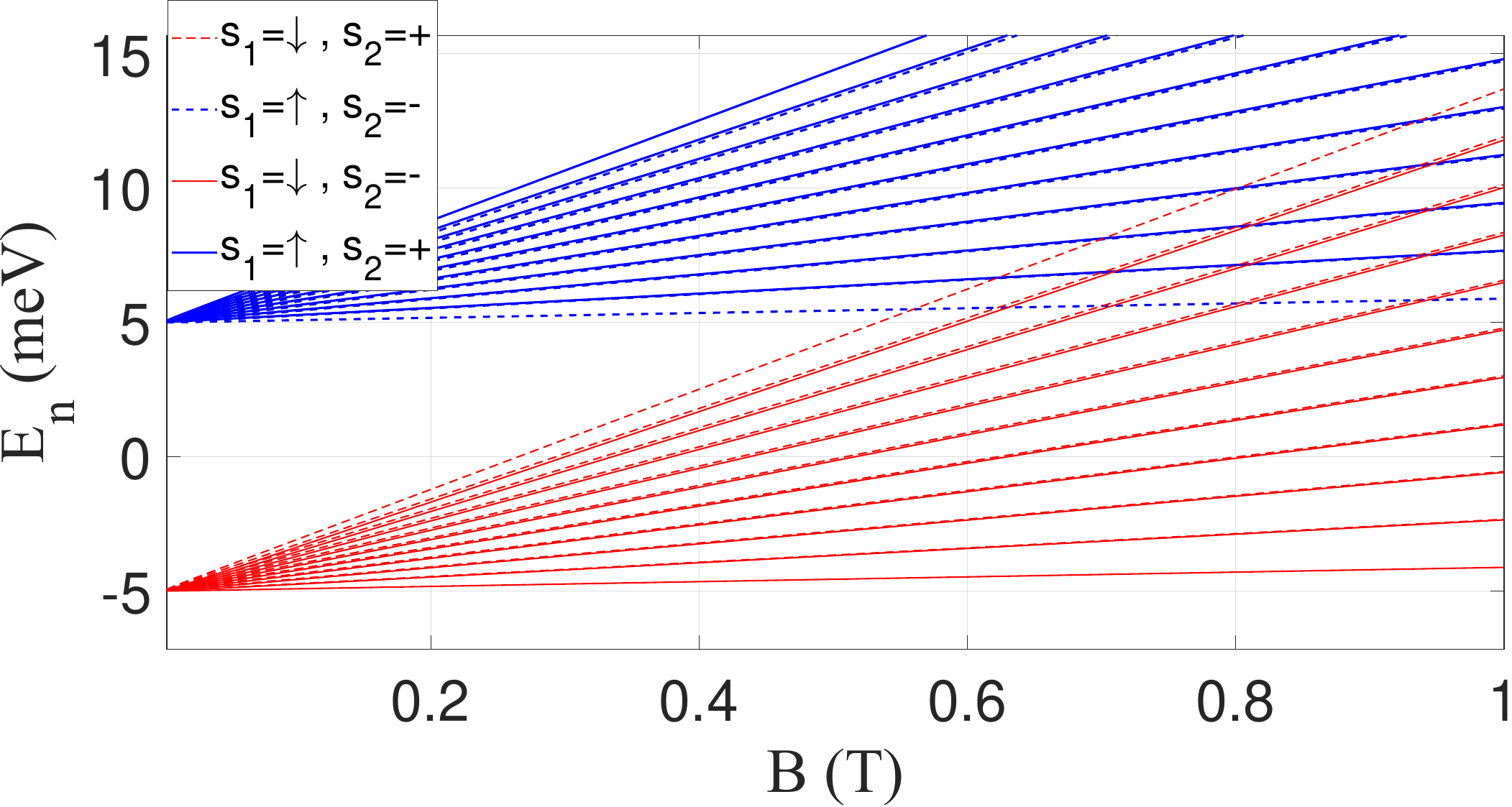}
         \caption{Few Landau levels are plotted with the magnetic field in both spin branches. Landau level crossings between opposite spin branches are clearly visible here. We use parameter as: $\eta = 5$~meV and $\alpha = 2 \times 10^{-12}$eV m.}
        \label{LL}
        \includegraphics[width=\linewidth]{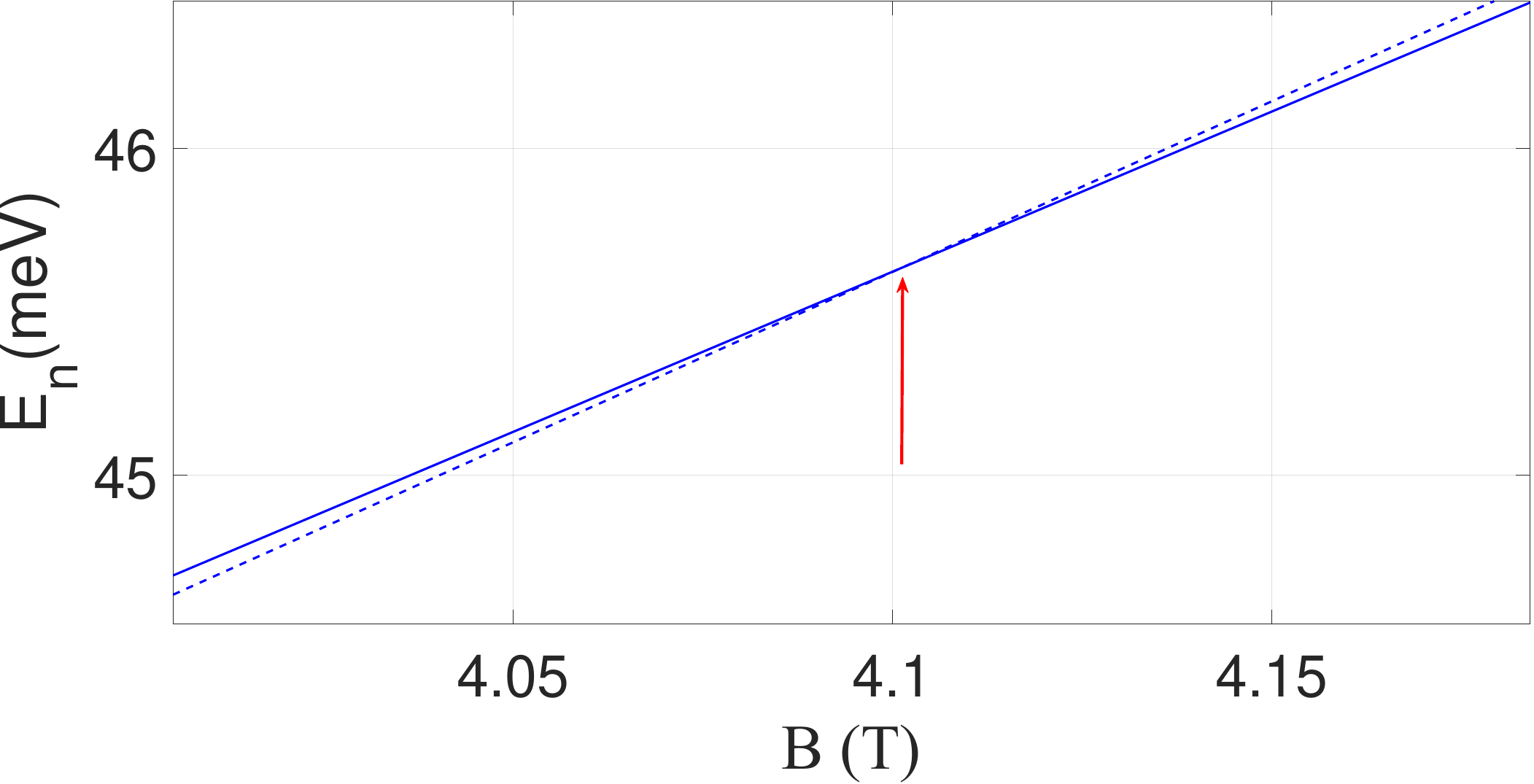}
        \caption{Crossing of two Landau levels(intra-spin inter-band) is shown at high magnetic field for $\eta = 5$~meV and $\alpha = 5 \times 10^{-12}$eV m. The crossing point is shown by an arrow(red)}
        \label{LL2}
    \end{minipage}
\end{figure}

 The Hamiltonian is translationally invariant along the $y$-direction as $[H,p_y]=0$, which allows
 us to write  the corresponding wave function as $\Psi(x,y)= e^{ik_yy}\phi(x)/\sqrt{L_y}$ where $L_y$ is the length along the free direction. Using this, the eigenvalue problem can be reduced to $H\phi(x)=E\phi(x)$ where
 \begin{widetext}
 \begin{equation}
H = \left[\begin{array}{cccc}
\hbar \omega_c(a^\dagger a +\frac{1}{2}) & -\beta a & -i\eta & -\beta a \\
-\beta a^\dagger & \hbar \omega_c(a^\dagger a +\frac{1}{2}) & \beta a^\dagger & i\eta \\
i\eta &  -\beta a   &  \hbar \omega_c(a^\dagger a +\frac{1}{2}) & -\beta a  \\
-\beta a^\dagger & -i\eta & -\beta a^\dagger &  \hbar \omega_c(a^\dagger a +\frac{1}{2})
\end{array}\right]
\end{equation}
\end{widetext}
 where $\omega_c=eB/m^*$ is the cyclotron frequency, \(\beta=\sqrt{2}\alpha/l_c\) is the energy scale associated with the RSOI term. Here, $l_c=\sqrt{\hbar/eB}$ is the magnetic length scale. The dimensionless ladder operators are defined as $a=(\tilde{x}+i\tilde{p})/\sqrt{2}$ and $a^{\dagger}=(\tilde{x}-i\tilde{p})/\sqrt{2}$. Here, the dimensionless position operator $\tilde{x}=(x+x_c)/l_c$ with centre of the cyclotron orbits at $x=-x_c-k_yl_c^2$, and the $x$-component of the momentum operator $\tilde{p}=-i\partial/\partial (x/l_c)$, satisfying the commutator relation \mbox{$[\tilde{x},\tilde{p}]=i\hbar$}. Note that we ignore here the Zeeman splitting term as it only renormalises the spin splitting energy. To solve the eigenvalue equation, we use the ansatz
\begin{equation}
\phi ({x}) =\sum_{\nu}\begin{bmatrix} a_{\nu} \\
b_{\nu} \\
c_{\nu}\\
d_{\nu}
\end{bmatrix}\phi_{\nu}({x}).
\end{equation}
where $\phi_{\nu}({x})$  is the one-dimensional harmonic oscillator wave function centred at $x_c$ and $\nu$ is an integer.  To proceed further, we multiply $\phi_{l}({x})$ by the eigenvalue equation from the left side and integrate over ${x}$, which yields the following four equations as 
\begin{widetext}
\begin{eqnarray}
&&\Bigl[\hbar \omega_c \!\left( l + \tfrac{1}{2} \right) -E \Bigr]\, a_{l}
- \beta \sqrt{l+1}\, b_{\,l+1}
- i \eta\, c_{l}- \beta \sqrt{l+1}\, d_{\,l+1} = 0,\label{ll1}\\
&-& \beta \sqrt{l}\, a_{\,l-1}
+ \Bigl[  \hbar \omega_c\!\left( l + \tfrac{1}{2} \right) - E \Bigr]\, b_{l}
- \beta \sqrt{l}\, c_{\,l-1}
+ i \eta\, d_{l} = 0\label{ll2},\\
&&i \eta\, a_{l}
- \beta \sqrt{l+1}\, b_{\,l+1}
+ \Bigl[  \hbar \omega_c \!\left( l + \tfrac{1}{2} \right) - E \Bigr]\, c_{l}- \beta \sqrt{l+1}\, d_{\,l+1} = 0,\label{ll3}\\
&-& \beta \sqrt{l}\, a_{\,l-1}
- i \eta\, b_{l}
- \beta \sqrt{l}\, c_{\,l-1}
+ \Bigl[  \hbar \omega_c \!\left( l + \tfrac{1}{2} \right) - E \Bigr]\, d_{l} = 0\label{ll4}.
\end{eqnarray}
\end{widetext}
 To diagonalize the above four equations, we perform $l\to l-1$ into Eq. (\ref{ll1}) and (\ref{ll3}), and switch to the new notation $n\equiv l-1$, which results in the Landau levels for ($n>0$) as  
\begin{equation}
E_{\zeta} = n \hbar \omega_c  + s_1\sqrt{A_n + s_2 B_n}
\label{LLS}
\end{equation}
where $\zeta \equiv \{n,k_y,s_1,s_2\}$ , $A_n = 2n\beta^2  +  (\hbar \omega_c/2)^2 +\eta^2$ and $B_n = \sqrt{ (\hbar \omega_c)^2\eta^2 + 4n^2\beta^4 + 4n\beta^2\eta^2 }$. 
The corresponding eigenstates are obtained as
\begin{equation}
\Psi_{n,k_y}^{s_1,s_2}(x,y)=\frac{e^{ik_y y}}{\sqrt{L_y}}
\begin{bmatrix}
a^{s_1,s_2}_{n} \phi_{n-1}(\tilde{x}) \\
b^{s_1,s_2}_{n}\phi_{n}(\tilde{x}) \\
c^{s_1,s_2}_{n}\phi_{n-1}(\tilde{x}) \\
d^{s_1,s_2}_{n}\phi_{n}(\tilde{x})
\end{bmatrix},
\end{equation}
where the coefficients $a^{s_1,s_2}_{n} $  $b^{s_1,s_2}_{n} $ $c^{s_1,s_2}_{n} $, and $d^{s_1,s_2}_{n} $ corresponding to each spinor are explicitly given in  Appendix (A).  The spinor component $\phi_n(\tilde{x})=[1/\sqrt{2^nn!l_c\sqrt{\pi}}]H_n(\tilde{x})\exp(-\tilde{x}^2/2)$ is the usual harmonic oscillator wave function.
The zero-th landau level is obtained separately by setting $l=0$ in Eq. (\ref{ll1}) and (\ref{ll3}) as $E_0 = (\alpha/2 )s_1\eta$ and the corresponding zero-th eigen state as
\begin{equation}
\Psi_{0,k_y}^{s_1}(x,y)=\frac{e^{ik_yy}}{\sqrt{L_y}}\left[\begin{array}{c}
                                  0\\
                                  b^{s_1,0}_{0}\phi_{0}(\tilde{x})\\
                                  0\\
                                   d^{s_1,0}_{0}\phi_{0}(\tilde{x})
                                  \end{array}\right],
\end{equation}
We comment here that the zeroth LL does not depend on the band index $s_2$, hence this is band degenerate. We now plot the LLs obtained in Eq.~(\ref{LLS})  as a function of magnetic field in Fig.~(\ref {LL}). All LLs linearly increase with magnetic field B that can be attributed to the first term of LLs in Eq.~(\ref{LLS}), i.e.,  $\hbar\omega_c\sim B$, the most dominating term. 
It is also observed that LL crossings between two bands or two spin branches occur while varying the magnetic field. We will show later that the effect of all these crossings can be clearly captured via quantum Hall quantization.  As we have seen that in the absence of a magnetic field, the spin splitting energy at $\Gamma$ point is solely determined by $\eta$ [see Eq.~(\ref{dispersion})], here also the spin splitting at $B\rightarrow 0$ is exclusively determined by $\eta$ only. The spin-splitting energy for a particular Landau level index $n$ is given by $\sqrt{A_n+s_2B_n}$, which also increases with the magnetic field $B$ and the LL index $n$. The inter-spin LL crossings are clearly visible in the Fig.~(\ref{LL}). However, two successive LLs belonging to two different bands with the same spin texture are very closely spaced, and for better visualisation  we plot $n=5$ and $6$ LLs for spin-up branch separately in the Fig.~(\ref {LL2}) where the crossing occurs at $B \approx 4.1$ T. Such crossing can also occur for higher LLs  within the same spin branch, but only at relatively high magnetic field. The energy spacing between two bands in the same spin branch is given by $ E_{n,s_1,+}-E_{n,s_1,-}=\sqrt{A_n + B_n}-\sqrt{A_n -B_n}$ that is dominated by $B_n$ which is in-fact function of both $\alpha$ and $\eta$ both, and increases with the magnetic field and LL index $n$
 
\subsection{Density of states}
\begin{figure}[H] 
    \centering
    \begin{subfigure}{0.47\textwidth}
        \centering
        \includegraphics[width=\linewidth]{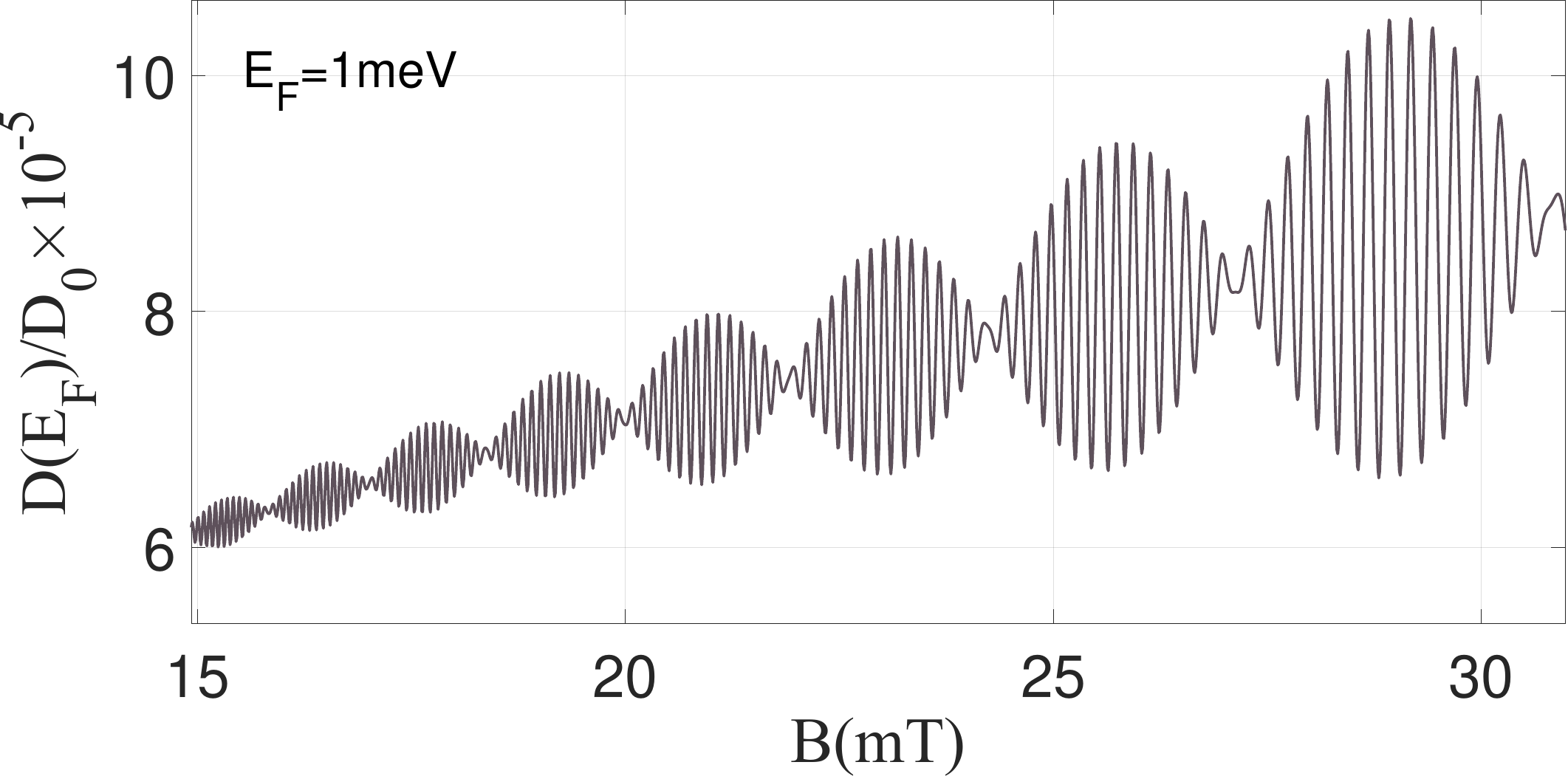}
        \caption{}
        \label{DOS_a}
    \end{subfigure}
    \hfill
    \begin{subfigure}{0.47\textwidth}
        \centering
        \includegraphics[width=\linewidth]{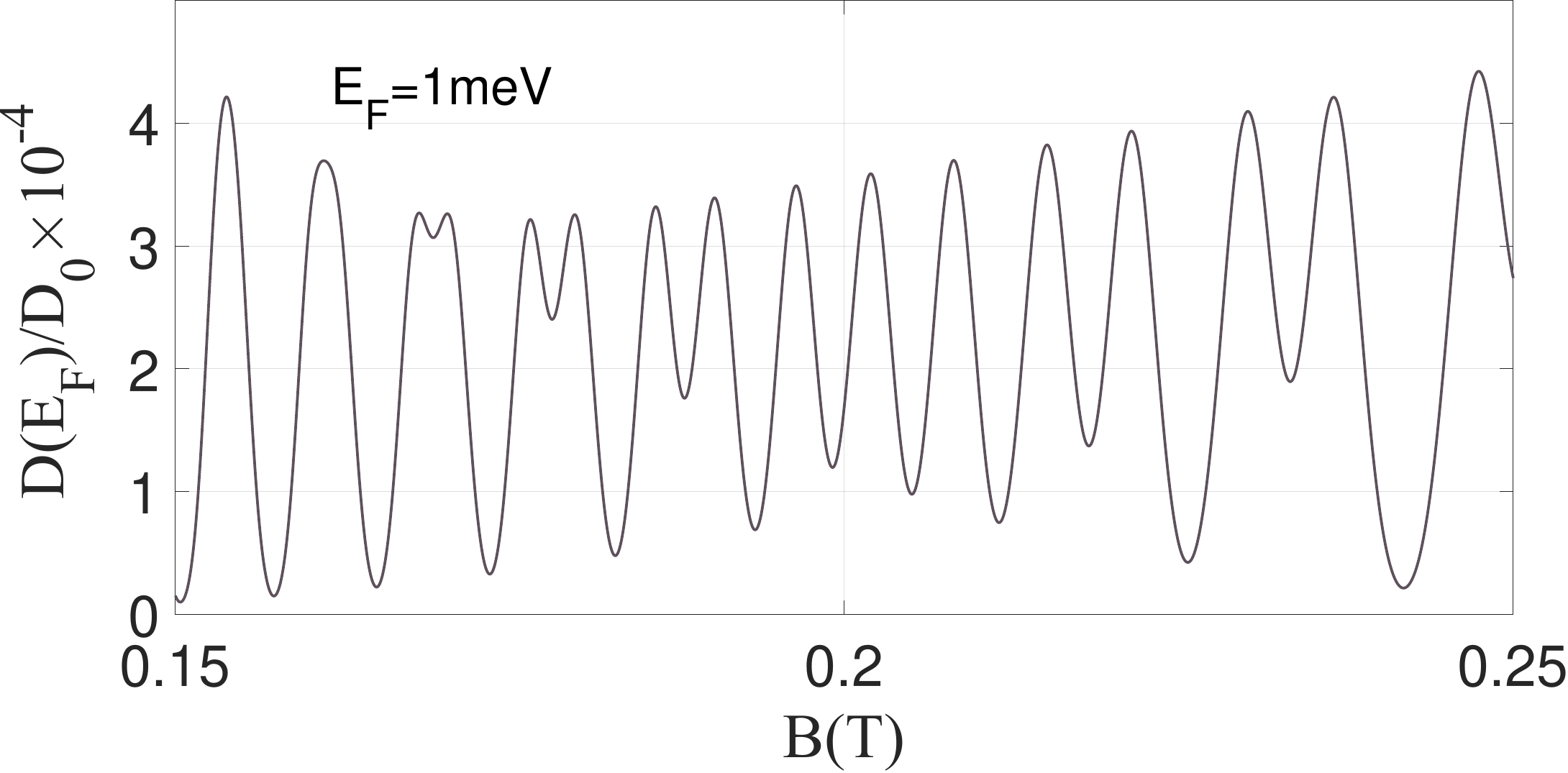}
        \caption{}
        \label{DOS_b}
    \end{subfigure}

   
    \begin{subfigure}{0.47\textwidth}
        \centering
        \includegraphics[width=\linewidth]{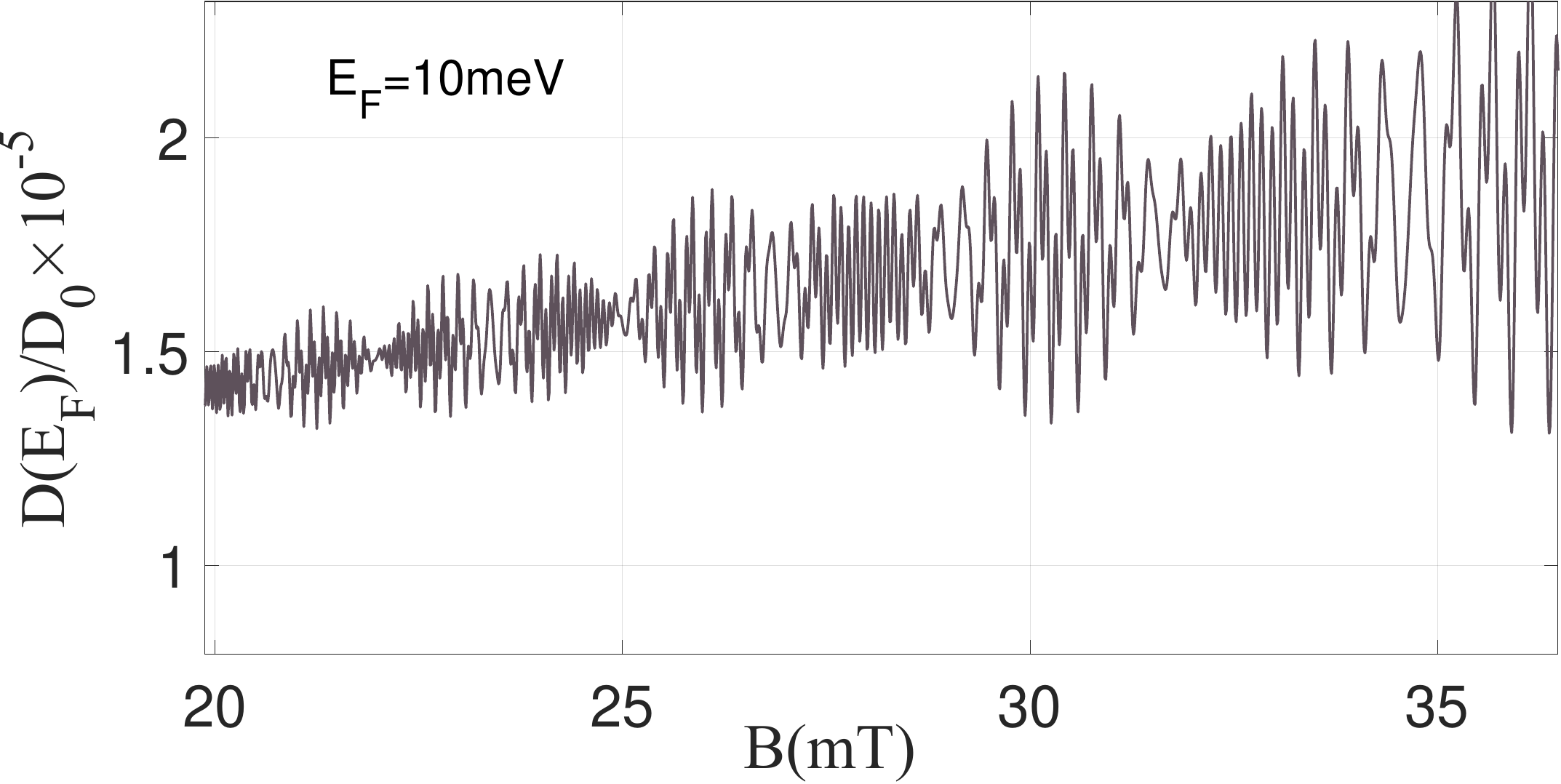}
        \caption{}
        \label{DOS_c}
    \end{subfigure}

    \caption{The DOS versus B are presented for different Fermi level and magnetic field range, such as (a) $E_F= 1$ meV at low magnetic field range, (b)  $E_F= 1$ meV but at relatively higher magnetic field (c)  $E_F= 10$ meV but at low magnetic field regime.}
        \label{DOS}
    \end{figure}

Now we briefly discuss the density of states (DOS) in the presence of a magnetic field. The DOS at energy $E$ can be expressed  as
\begin{equation}\label{delta}
D(E)=\frac{1}{\Omega}\sum_{\zeta}\delta(E-E_{\zeta})
\end{equation}
where $\Omega=L_x\times L_y$. To perform the summation over $k_y$ we use the fact that the center of the cyclotron orbits $x_c=k_yl_c^2$ is always restricted to the system size i.e., $0<k_yl_c^2<L_x$ which yields $\sum_{k_y} \;\rightarrow\; (L_y / 2\pi) \int_0^{L_x / l_c^2} dk_y \;=\; \Omega / (2 \pi l_c^2)$ which is the degeneracy of each Landau level. Using this, we get
\begin{equation}\label{dos}
 D(E)= \frac{1}{2\pi l_c^2}\sum_{n,s_1,s_2}\delta(E-E_{n,s_1,s_2})
\end{equation}
which exhibits a series of delta functions with energy. However, it's valid for a clean system without any impurities or disorder. In a realistic situation, the LLs are always broadened due to the presence of impurities. The impurity-induced Landau level broadening can be incorporated via a Gaussian distribution of the delta function as
\begin{equation}\label{dos1}
 D(E)= D_0\sum_{n,s_1,s_2}\exp\left[-\frac{(E-E_{n,s_1,s_2})^2}{2\Gamma^2}\right]
\end{equation}
where $D_0 = 1 / (2 \pi l_c^2 \, \Gamma \sqrt{2 \pi})$ and  is the width of Gaussian distribution.  The impurity-induced LL broadening was experimentally found to be weakly dependent on the magnetic field as $\Gamma\propto \sqrt{B}$. We keep broadening weak as $\Gamma=0.1\sqrt{B}$ meV to ensure that LLs are well separated. We plot DOS at $E=E_F$ using Eq.~(\ref{dos1}) in Fig. (\ref {DOS}) showing oscillations with the magnetic field. This oscillation is a direct manifestation of discrete energy levels (LLs), also known as the Shubnikov-de Haas oscillation.

 We also note in  Fig.~(\ref{DOS_a}) the appearance of a beating pattern when the Fermi level is placed at $E_F=1$ meV. This can be attributed to the superposition of two oscillatory DOS with closely spaced frequencies, belonging to two bands of the same spin branch. Here, we comment that such beating was also noted in the conventional RSOI system, but there it was due to the superposition of two opposite spin branches \cite{Das1989, Wang2003}.
This beating pattern persists up to $B \approx 0.16$~T, as illustrated in Fig.~(\ref{DOS_b}). In the higher magnetic field regime, however, the beating is suppressed due to the significant frequency difference between the two bands. In Fig.~(\ref{DOS_c}), the Fermi level is set relatively high, allowing contribution to the DOS from both branches, which leads to a distorted beating pattern.  
\begin{figure}[H] 
    \centering
  \begin{subfigure}{0.47\textwidth}
        \centering
        \includegraphics[width=\linewidth]{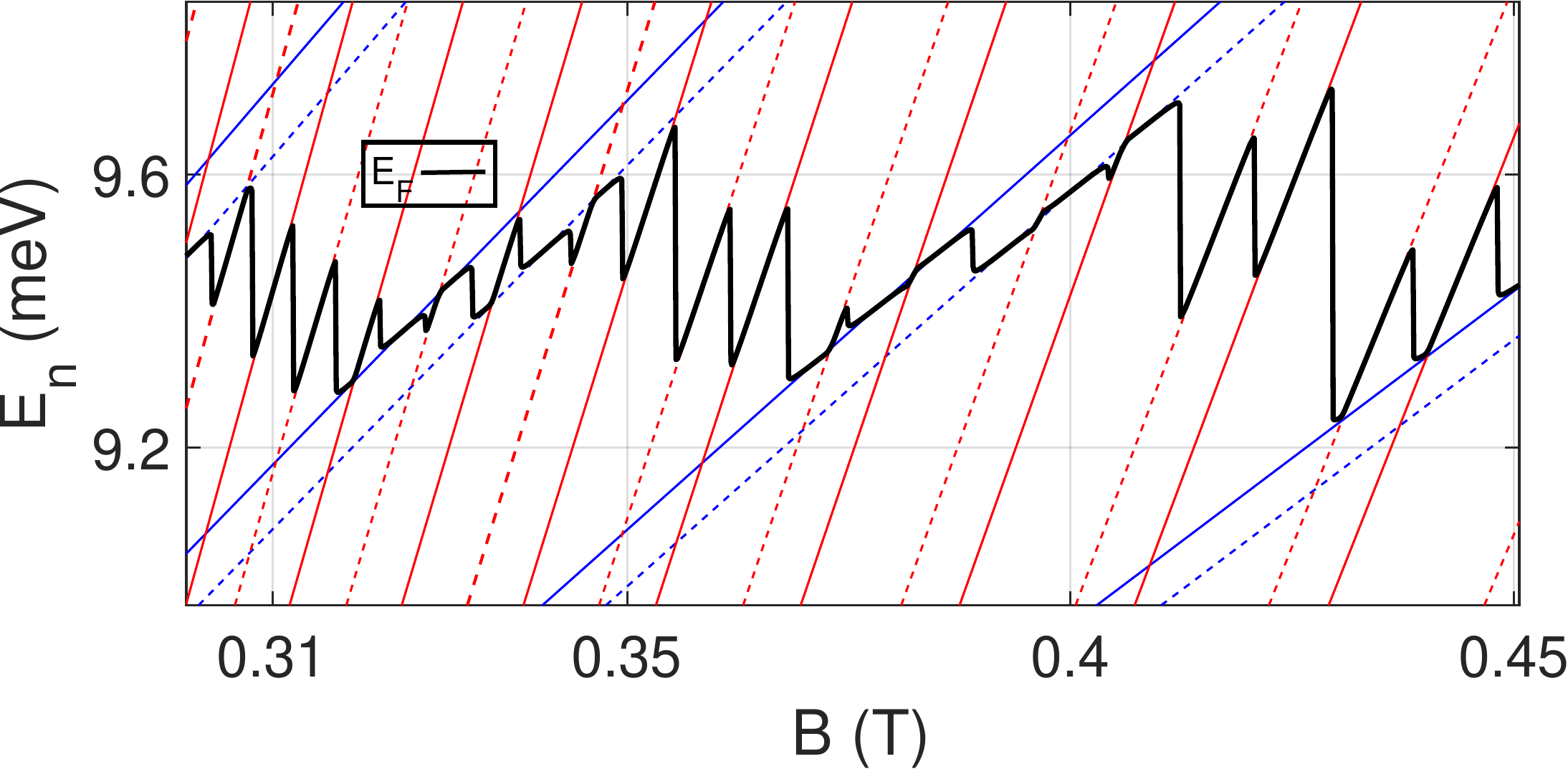}
        \caption{}
        \label{}
    \end{subfigure}
    \hfill
    \begin{subfigure}{0.47\textwidth}
        \centering
        \includegraphics[width=\linewidth]{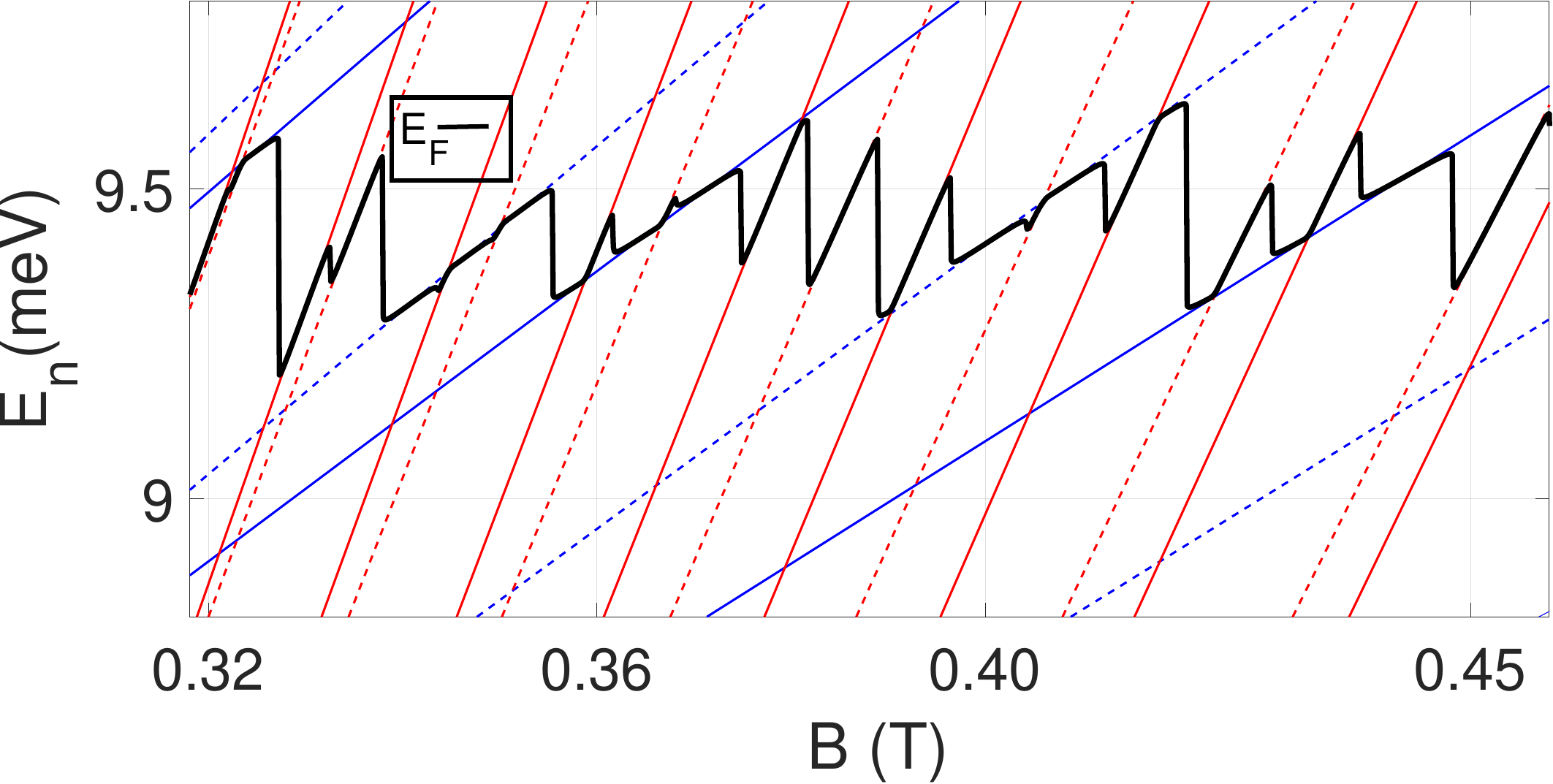}
        \caption{}
        \label{}
    \end{subfigure}
    \caption{Landau levels and Fermi level of the system at $T=5$mK as a function of magnetic field is shown for (a) $\alpha=7\times10^{-12}$eVm and (b) $\alpha=5\times10^{-12}$eVm. Carrier concentration is taken to be $n_e = 2 \times 10^{16} \, m^{-2}$}
    \label{Fermilevel}
\end{figure}
Now we briefly discuss the behaviour of the Fermi level with the magnetic field for a fixed carrier density. To obtain the Fermi energy ($E_F$) in the presence of a magnetic field, we start with the expression of carrier density $n_e$ as
\begin{equation}
n_e = \int_{-\infty}^{+\infty} D(E) f(E) \, dE
\end{equation}
where $D(E)$ is the DOS given by Eq.~(\ref{dos}). The Fermi-Dirac distribution function is given by $f(E) = \left[\exp\bigl\{(E-E_F)/k_B T\bigr\} + 1\right]^{-1}$. Now using Eq.~(\ref{dos}) we obtain the self-consistent equation
\begin{equation}
2\pi n_e \, l_c^2 = \sum_{n,s_1,s_2} f(E_{n,s_1,s_2})
\end{equation}
that is solved numerically to obtain the Fermi level as a function of magnetic field, and shown in Fig.~(\ref{Fermilevel})

The Fermi level is found to be weakly oscillating with the magnetic field among different Landau levels, attributed to different spin branches and bands. We show here two plots for two different strengths of RSOI, and this does not affect the qualitative feature of the Fermi level as RSOI enters through spin splitting of LLs.

\section{Magnetotransport Properties}\label{longandhall}

In this section, we study the magnetotransport properties.  We adopt the formalism developed in Ref.~\cite{Charbonneau1982}, and frequently used in Refs.~\cite{Wang2003, Islam2018,Biswas2016,Sinha2020,Tahir2013, krstajic2012}. To incorporate responses to a time-dependent external AC bias, the many-body Hamiltonian can be written as $\mathcal{H}=H+H_I-e\mathbf{R.E}(t)$. Here, $H$ is the unperturbed Hamiltonian, $H_I$ describes interactions between electrons and impurities or phonons, $\mathbf{E}(t)$ is the external AC bias, and $\mathbf{R=\sum_{r_i}}$ with $\mathbf{r_i}$ being the position vector of the $i$-th electron. In one electron approximation the response of the electronic system, in presence of a perpendicular magnetic field, to the applied bias can be described by the conductivity tensor: $\sigma_{\mu \nu}=\sigma^{d}_{\mu \nu}+\sigma^{nd}_{\mu \nu}$, with $\mu,\nu=x,y$. Here, the diagonal component $\sigma^d_{\mu\nu}$ gives the longitudinal conductivity, whereas the off-diagonal component $\sigma^{nd}_{\mu\nu}$ gives Hall conductivity. These transport coefficients ($\sigma^d_{\mu\nu}$ and $\sigma^{nd}_{\mu\nu}$) have been evaluated in Ref.~\cite{Charbonneau1982}.

\subsection{Longitudinal conductivity} In this subsection, we calculate the longitudinal conductivity that mainly arises from the scattering of the cyclotron orbits with charge impurities. To proceed further, we consider that the scattering process is fully elastic in nature in the low temperature regime. It can be justified from the fact that the charge carriers cannot supply enough energy to excite the heavy ionised impurity from its ground state to exited state. Apart from the scattering mechanism, non-zero group velocity can also contribute to the transport process. In this context, the longitudinal conductivity generally comprises of two primary components: the diffusive and collisional contributions, i.e., the diagonal part of the conductivity tensor can be written as $\sigma^{d}_{\mu\nu}=\sigma^{\rm diff}_{\mu\nu}+\sigma^{\rm coll}_{\mu\nu}$. Here, the diffusive contribution can be obtained as
\begin{equation}
\sigma^{\rm diff}_{\mu\nu}=\frac{\beta_T e^2}{2\Omega }\sum_{\zeta,\zeta'}f_{\zeta}(1-f_{\zeta'})v_{\mu}v_{\nu}.
\end{equation}
 Here, $\Omega$ is the area of the sample, $T$ is the temperature, and the Fermi distribution function is given by $f_\zeta = \left[\exp\bigl\{(E_\zeta-E_F)/k_B T\bigr\} + 1\right]^{-1}$,  the group velocity $v_\mu=\hbar^{-1}\partial E_\zeta/\partial k_\mu$ which is zero in the present case as all the LLs are disperssionless flat causing strong localization to the cyclotron orbits, hence no diffusion contribution to the longitudinal conductivity (also known as diffusive conductivity) appears.
 
We now consider the collisional conductivity that is given by  \cite{Charbonneau1982}
\begin{equation}
\sigma^{coll}_{xx}=\frac{\beta_T e^2}{2\Omega }\sum_{\zeta,\zeta'}f_{\zeta}(1-f_{\zeta'})
W_{\zeta,\zeta'}(x_{\zeta}-x_{\zeta'})^2.
\end{equation}
where the average of the $x$ component of the position operator in the state $\ket{\zeta}$ is $x_\zeta = \langle \zeta | x | \zeta \rangle = k_y l_c^2$. Hence, $(x_\zeta-x'_\zeta)^2=(q_yl_c^2)^2$ where $q_y=k_y'-k_y$ is the momentum transfer between two states before and after scattering. Moreover, $W_{\zeta \zeta'}$ describes the probability that an electron is scattered from an initial state $\ket{\zeta}$ to a final state $\ket{\zeta'}$, and its probability can be described by the usual Fermi-Golden rule as
\begin{equation}
W_{\zeta, \zeta'} = \frac{2 \pi n_{\mathrm{im}}}{\hbar \Omega} 
\sum_{\mathbf{q}} \lvert U(\mathbf{q}) \rvert^2 \lvert F_{\zeta, \zeta'} \rvert^2 
\delta(E_{\zeta} - E_{\zeta'}) \, \delta_{k_y , k_y' +q_y}.
\label{scattering}
\end{equation}
Here, $n_{\text{im}}$ is the impurity density, and $U(\mathbf{q})$ is the Fourier transform of the screened Coulomb potential 
 $U(\mathbf{r}) = (e^2 e^{-k_s r}) / (4 \pi \epsilon_0 \epsilon_r r)$
where $\epsilon_0$ is the free space permittivity, $\epsilon_r$ is the dielectric constant of the medium, and $k_s$ is the screened wave vector. 
For a 2D system, its Fourier transform is given by \( U(\mathbf{q}) = e^2 / \bigl( 4 \pi \epsilon_0 \epsilon_r \sqrt{q^2 + k_s^2} \bigr) \).
The form factor $F_{\zeta,\zeta'}= \langle \zeta' | e^{i \mathbf{k} \cdot \mathbf{r}} | \zeta \rangle$ and $|F_{\zeta,\zeta'}|^2$ describes the probability of scattering an electron from a quantum state $|\zeta\rangle$ to $|\zeta'\rangle$. We restrict ourselves to intra-branch ($s_1=s_1'$) intra-band ($s_2=s_2'$) and intra-level ($n' = n$) scattering because of the presence of the term $\delta(E_{\zeta} - E_{\zeta'})$ in Eq.~\eqref{scattering} preserving the elastic nature of scattering. 
The form factor is simplified as 
\begin{equation}
\begin{aligned}
|F|^{n}_{s_1,s_2} &= \lvert a^{n}_{s_1,s_2} \rvert^2 L_{n-1}(u) 
+ \lvert c^{n}_{s_1,s_2} \rvert^2 L_{n-1}(u) \\[6pt]
&\quad + \lvert b^{n}_{s_1,s_2} \rvert^2 L_{n}(u) 
+ \lvert d^{n}_{s_1,s_2} \rvert^2 L_{n}(u)
\end{aligned}
\end{equation}

The sharp Landau levels are broadened by the static impurities present in the system, and the delta function can be replaced by Lorentzian broadening, represented as 
$\delta(E_{\zeta} - E_{\zeta'}) = \pi \Gamma / [(E_{\zeta} - E_{\zeta'})^2 + \Gamma^2]$ where $\Gamma$ is the Landau level broadening. 
For intra-level and intra-band scattering, this may be approximated as $\delta(E_{\zeta} - E_{\zeta'}) \simeq \frac{1}{\pi \Gamma}$.
Further, $U(\mathbf{q})$ is approximated as \( U(\mathbf{q}) \simeq e^2 / (4 \pi \epsilon_0 \epsilon_r k_s) \equiv U_0 \),
since small values of $q^2$ contribute more due to the exponentially decaying term $e^{-u}$ in the expressions of $|I^{n}_{s_1,s_2}|^2$. 
Using the facts that $\sum_{k_y} \;\rightarrow\; 1 / (2 \pi l_c^2)$  and $\sum_{\mathbf{q}} \rightarrow \Omega / (2\pi)^2 \int q \, dq \, d\theta$
where $\theta$ is the polar angle of $\mathbf{q}$, we finally obtain the following expression for the longitudinal conductivity as ...
\begin{equation}\label{col}
\sigma_{xx} = \tilde{\sigma}_0  \sum_{n,s_1,s_2}I^{n}_{s_1,s_2} 
f\bigl(E_{n,s_1,s_2}\bigr) 
\left[ 1 - f\bigl(E_{n,s_1,s_2}\bigr) \right].
\end{equation}
Here, 
$\tilde{\sigma}_0 = (e^2 \beta n_{\text{im}} U_0^2) / (\pi \hbar \Gamma l_c^2)$,
and 
$I^{n}_{s_1,s_2} = \int_0^{\infty} u \, |F^{n}_{s_1,s_2}(u)|^2 \, du$. Using the standard results for Laguerre polynomials 
$\int_0^{\infty} L_n^2(u) e^{-u} u \, du = 2n+1$ and $\int_0^{\infty} L_n(u) L_{n-1}(u) e^{-u} u \, du = -n$, we obtain
\begin{equation}
\begin{aligned}
I^{n}_{s_1,s_2} &= (2n - 1)\,\big(\lvert a^{n}_{s_1,s_2} \rvert^2  + \lvert c^{n}_{s_1,s_2} \rvert^2 \big)^2  \\
&\quad + (2n + 1)\,\big(\lvert b^{n}_{s_1,s_2} \rvert^2  + \lvert d^{n}_{s_1,s_2} \rvert^2 \big)^2  \\
&\quad - 2n\,\big(\lvert a^{n}_{s_1,s_2} \rvert^2  + \lvert c^{n}_{s_1,s_2} \rvert^2 \big)\big(\lvert b^{n}_{s_1,s_2} \rvert^2  + \lvert d^{n}_{s_1,s_2} \rvert^2 \big)
\end{aligned}
\end{equation}
\begin{figure}[H]
    \centering 
    
    \begin{subfigure}{0.5\textwidth}
        \centering
        \includegraphics[width=\linewidth]{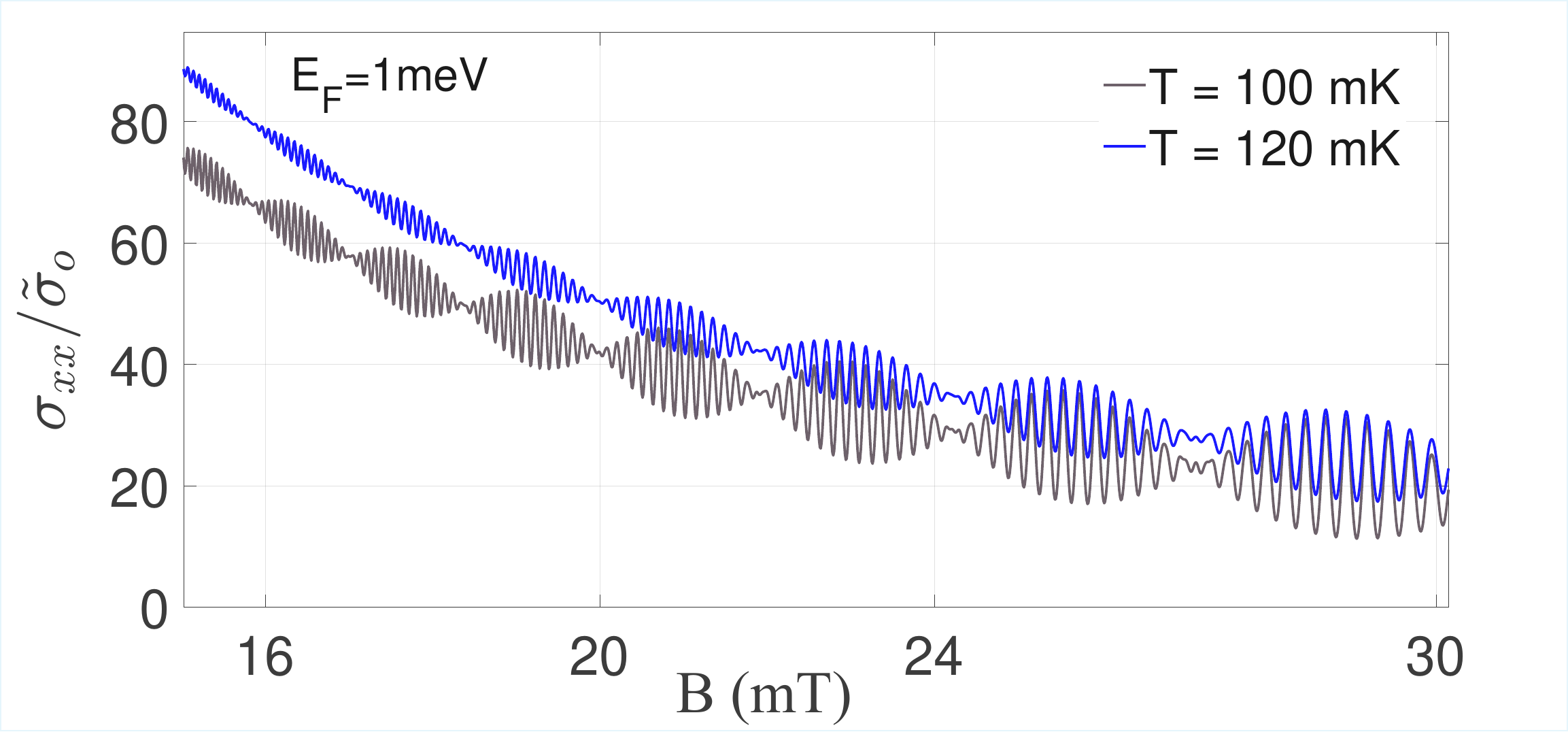}
        \caption{}
        \label{col_a}
    \end{subfigure}
    
    \begin{subfigure}{0.47\textwidth}
        \centering
        \includegraphics[width=\linewidth]{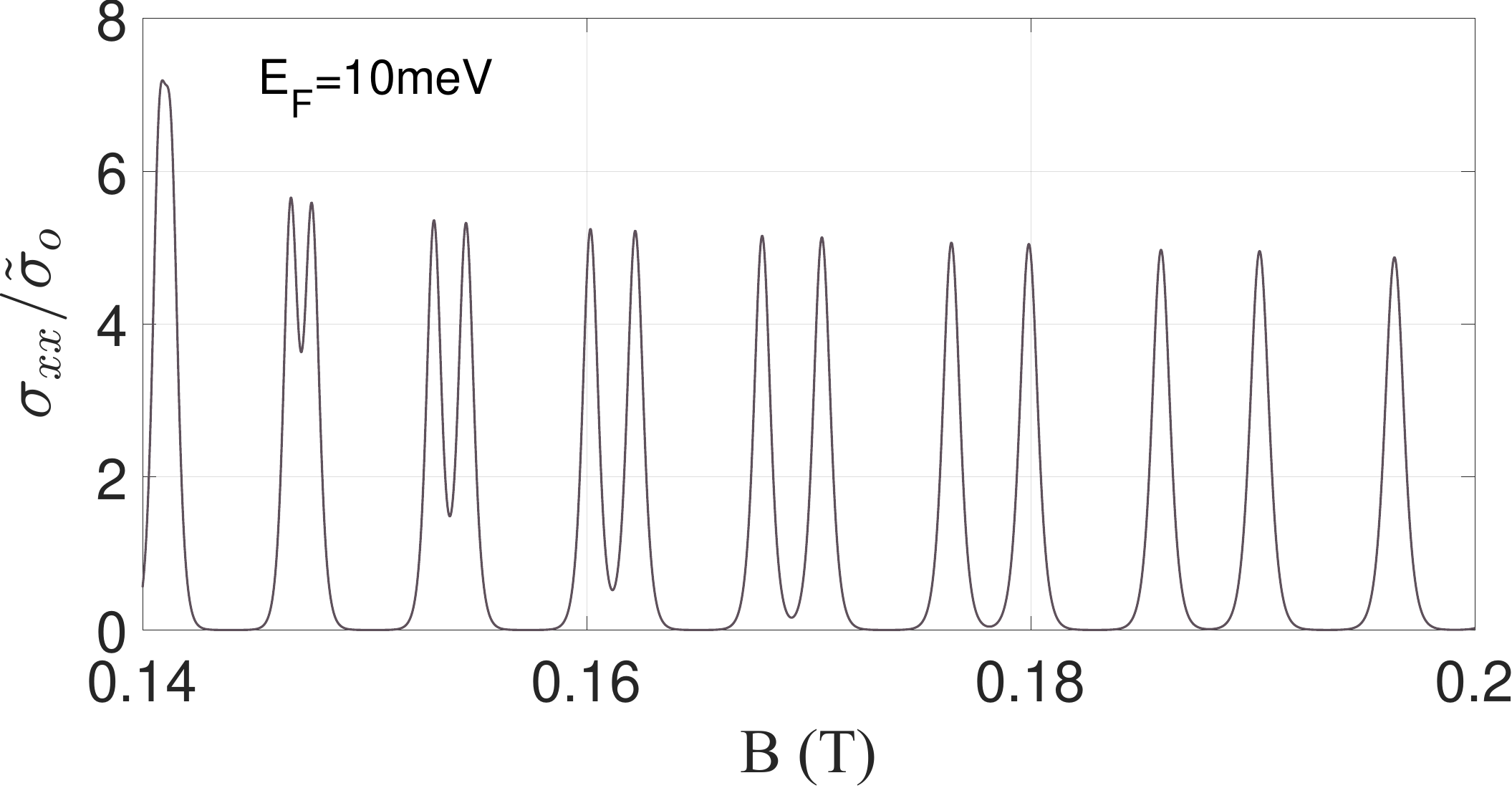}
        \caption{}
        \label{col_b}
    \end{subfigure}

    \begin{subfigure}{0.47\textwidth}
        \centering
        \includegraphics[width=\linewidth]{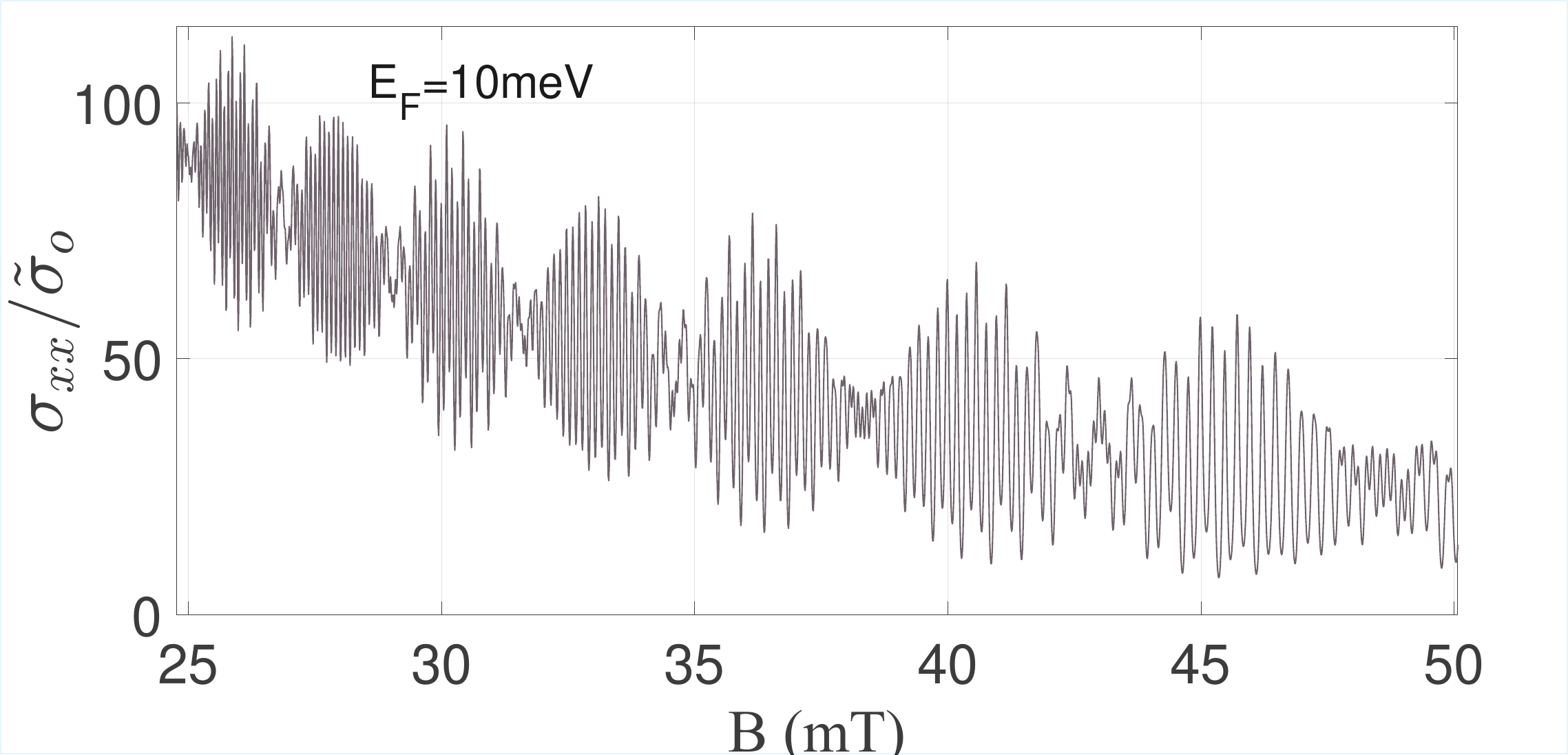}
        \caption{}
        \label{col_c}
    \end{subfigure}

    \caption{The longitudinal conductivity versus magnetic field are depicted in Fig. \ref{col_a}–\ref{col_c}. The Fig. (\ref{col_a}) presents the results for $E_F = 1$ meV in the low magnetic field regime at two different temperatures, revealing a beating pattern arising from the superposition of two closely spaced SdH oscillations within the spin-down branch. In the Fig. (\ref{col_b}), corresponding to a relatively higher magnetic field. Whereas the Fig. (\ref{col_c}) shows the case for a higher Fermi energy $E_F = 10$~meV occupying both spin branches, showing a distorted beating pattern in SdH oscillation.}
   \label{collision}
\end{figure}

We compute the collisional conductivity using the Eq.~(\ref{col}) and plot it with the magnetic field in Fig.~(\ref{collision}). We first plot the longitudinal conductivity in the low magnetic field regime at $E_F = 1$ meV in Fig.~(\ref{col_a}), occupying only the spin-down branch. In this case, we observe a beating pattern which is due to the superposition of SdH oscillations with closely separated frequencies attributed to two bands in the spin-down branch. Here, we also use two different temperatures, showing the suppression of SdH oscillation. The beating pattern phenomena in SdH oscillation in the longitudinal conductivity are also a direct manifestation of similar phenomena in the oscillatory DOS. The appearance of a beating pattern is direct evidence of the presence of two closely spaced bands.  We plot the same for a relatively high magnetic field regime in Fig.~(\ref{col_b}), which shows that the beating pattern vanishes, that is because the frequency difference between the SdH oscillations corresponding to the two bands also gets widened. 

Now we plot the SdH oscillation by keeping $E_F = 10$~meV, occupying both the spin-branches in the Fig.~\eqref{col_c}. We observe here that because of the superposition of SdH oscillations coming from two spin-branches, each having two bands, the beating pattern gets distorted, although the SdH oscillation remains intact. Here, we can clearly see that by keeping the Fermi level suitably, we can achieve a fully spin-polarized longitudinal conductivity as two opposite spin branches are well separated in energy space. Finally, we comment here that in a conventional 2D electronic system with conventional RSOI, such a beating pattern was observed, confirming the existence of two spin-branches. Contrary to that, in the present case, the beating pattern is due to the superposition of the SdH oscillations coming from the two bands in each spin branch \cite{Wang2005}.

\subsection{Quantum Hall conductivity}
\begin{figure}[H]
 \hspace{0cm}
\includegraphics[width=0.49\textwidth]{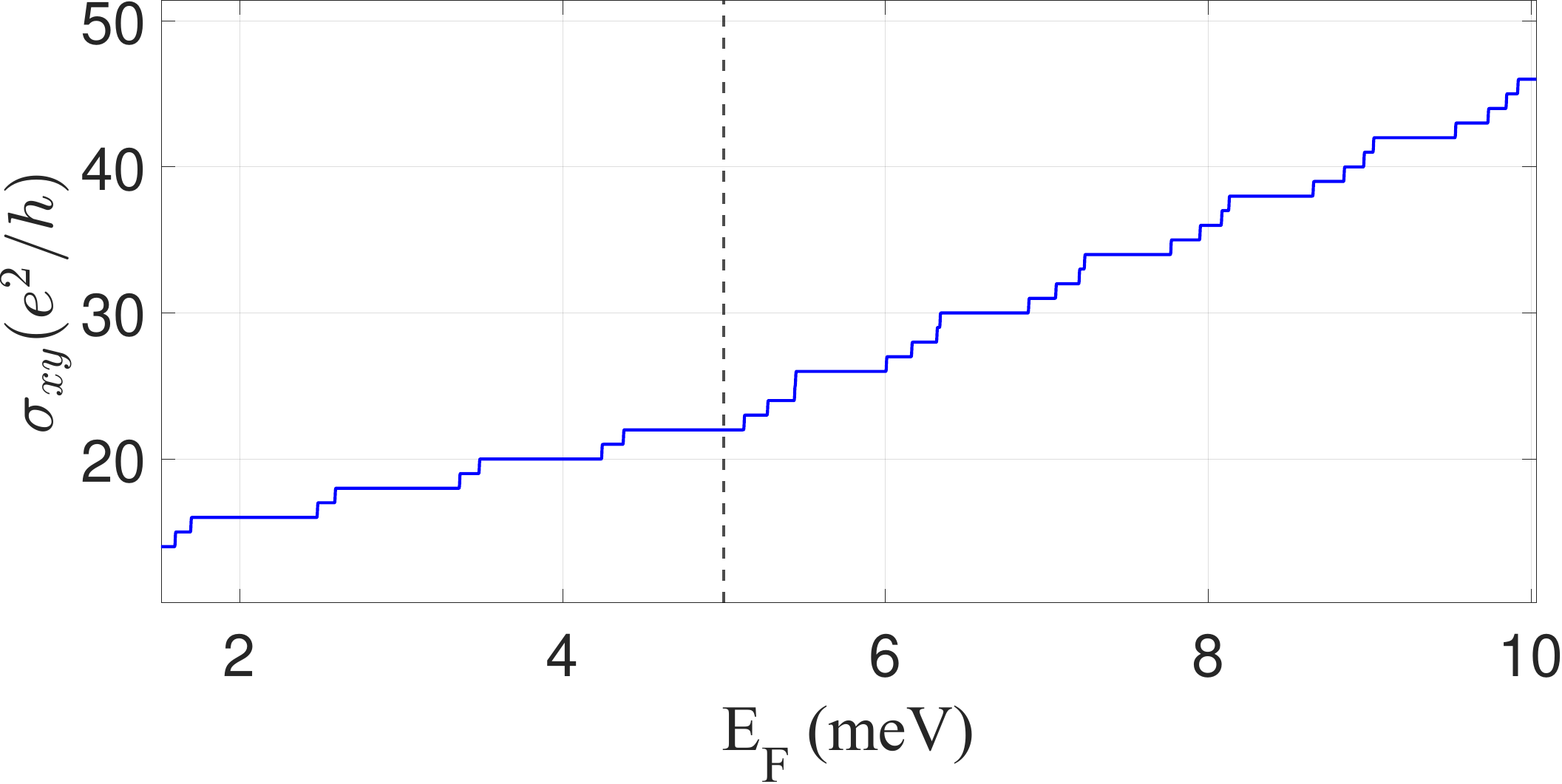}
\caption{The Hall conductivity versus Fermi energy at $B=0.5$ T and temperature $T=0.01$ K.}
  \label{Hall_Ef}
\end{figure}

In this section, we evaluate  the Hall conductivity aiming to examine the usual quantization and most importantly the impact of LL crossing. It's general expression is given as  \cite{Charbonneau1982,Wang2003,Tahir2016,Biswas2016}
\begin{equation}
\sigma_{xy}=\frac{ie^2\hbar}{\Omega}\sum_{\zeta\ne\zeta'}(f_{\zeta}-f_{\zeta'})
\frac{\langle\zeta| \mathcal{\hat{V}}_x| \zeta'\rangle\langle\zeta'|\mathcal{\hat{V}}_y|\zeta\rangle}
{(E_{\zeta}-E_{\zeta'})(E_{\zeta}-E_{\zeta'}+i\Gamma_0)}.
\label{hall conductivity}
\end{equation}
where \( f_{\zeta} \equiv f(E_{\zeta}) \) . The expression for velocity matrix can be computed using $\hat{\mathcal{V}}_x = \partial H / \partial p_x$ and $\hat{\mathcal{V}}_y = \partial H / \partial p_y$
\begin{align}
    \hat{\mathcal{V}}_x &= \frac{p_x}{m^\ast} \tau_0  \sigma_0 + \frac{\alpha}{\hbar} (\tau_0 + \tau_1)  \sigma_y \label{eq:vx} \\
    \hat{\mathcal{V}}_y &= \frac{(p_y+eBx)}{m^\ast}\;\tau_0  \sigma_0- \frac{\alpha}{\hbar} (\tau_0 + \tau_1)\sigma_x \label{eq:vy}
\end{align}
Now, the summation in Eq.~\eqref{hall conductivity} can be simplified as 
$\sum_{\zeta,\zeta'} \;\longrightarrow\; \frac{\Omega}{2\pi l_c^2}\sum_{s_1,s_1',s_2,s_2'}\sum_{n,n'}$. 
Assuming the broadening of Landau levels is the same for all states, the imaginary part vanishes, $\Gamma_0 = 0$. 
Therefore, Eq.~\eqref{hall conductivity} becomes 
\begin{equation}\label{hall_formula}
\begin{aligned}
\sigma_{yx} = & \frac{i \hbar e^2}{2 \pi l_c^2} 
\sum_{\substack{s_1, s_2, s_1', s_2' \\ n \ne n'}} 
\frac{(f^{s_1, s_2}_{n} - f^{s_1', s_2'}_{n'})}
{(E_{n, s_1, s_2} - E_{n', s_1', s_2'})^2} \\
& \times Q_{nn'}^{s_1, s_1', s_2, s_2'} 
\, \delta_{s_1, s_1'} \delta_{s_2, s_2'}
\end{aligned}
\end{equation}
where \(Q_{nn'}^{s_1,s_1',s_2,s_2'} = \langle \Psi_{nk_y}^{s_1, s_2} | \mathcal{V}_x | \Psi_{n',k_y'}^{s_1',s_2'} \rangle \langle \Psi_{n',k_y'}^{s_1',s_2'} | \mathcal{V}_y | \Psi_{n,k_y}^{s_1,s_2} \rangle\)
The velocity matrix is evaluated as
\begin{align}
Q_{nn'}^{s_1,s_1',s_2,s_2'} &=\Big( A_{nn'}^{s_1,s_1',s_2,s_2'} \, \delta_{n,\,n'+1} + B_{nn'}^{s_1,s_1',s_2,s_2'} \, \delta_{n,\,n'-1} \Big) \nonumber \\[6pt]
&\quad \times
\Big( C_{nn'}^{s_1,s_1',s_2,s_2'} \, \delta_{n,\,n'+1} + D_{nn'}^{s_1,s_1',s_2,s_2'} \, \delta_{n,\,n'} \Big)
\end{align}
where the expression for \(  A_{nn'}^{s_1,s_1',s_2,s_2'}, B_{nn'}^{s_1,s_1',s_2,s_2'}, C_{nn'}^{s_1,s_1',s_2,s_2'}, D_{nn'}^{s_1,s_1',s_2,s_2'}\) are quite lengthy and given in Appendix (B).
The zeroth LL contribution has to be evaluated separately as 
\begin{align*}
\sigma^{(0)}_{yx} &= \frac{i \hbar e^2}{2 \pi l_c^2}\sum_{s_1,s_1's_2,s_2'} \Bigg[
\frac{(f_{1}^{s_1,s_2} - f_{0}^{\ s_1',s_2'}) \, Q_{10}^{s_1,s_1',s_2,s_2'}}
     {(E_{1,s_1,s_2} - E_{0,s_1',s_2'})^2} \\
&\quad + \frac{(f_{0}^{s_1,s_2} - f_{1}^{s_1',s_2'}) \, Q_{01}^{s_1,s_1',s_2,s_2'}}
       {(E_{0,s_1,s_2} - E_{1,s_1',s_2'})^2}
\Bigg] \delta_{s_1,s_1'} \delta_{s_2,s_2'}
\end{align*}
The matrix product \( Q_{10} \) and \( Q_{01} \) are given in Appendix (B). We first plot the quantum-Hall conductivity with the Fermi level by using Eq.~(\ref{hall_formula}), shown in Fig.~(\ref{Hall_Ef}). In the low Fermi energy regime (upto $5$meV), only the LLs corresponding to the spin-down branch are occupied, and contribute to the Hall conductivity. 
The usual Hall quantization in units of $e^2/h$ corresponds to each Landau level appears. However, we note that as the Fermi level passes through $5$ meV (indicated by a dashed line), the Hall plateaus becomes smaller and the closely spaced Hall steps increase. This is because the Fermi level is now occupying both the spin branches, and the Landau level spacing between two successive LLs has now been decreased, which can also be seen from the Fig.~(\ref{LL}). Once the Fermi level occupies both spin branches, closely spaced four successive Hall steps also start emerging. Note that although the Fermi level weakly fluctuates among different LLs as the magnetic field is varied at fixed carrier density, here we keep the Fermi level constant, assuming that the fluctuations are very weak or that the carrier density is adjusted by gate voltage to keep the Fermi level constant as the magnetic field is tuned.

\begin{figure}[H]
 \hspace{0cm}
\includegraphics[width=0.49\textwidth]{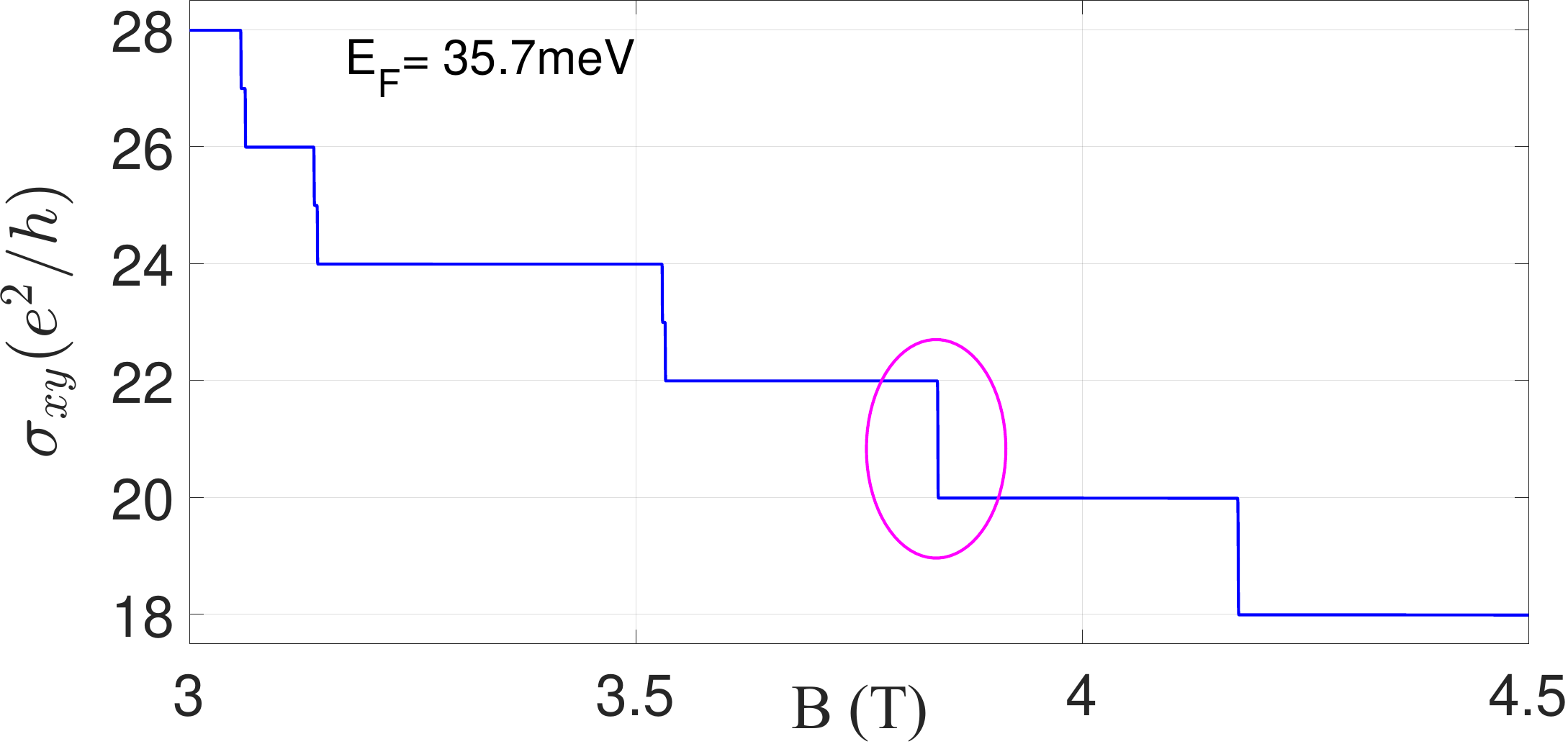}
\caption{The Hall conductivity versu magnetic field is plotted showing double jump corresponds to the inter-spin branch LLs crossing. The double jump in the Hall quantization is marked with an oval. The temperature is kept very low, at $T=10$ mK.}
 \label{H1}
\end{figure}

\begin{figure}[H]
 \hspace{0cm}
\includegraphics[width=0.49\textwidth]{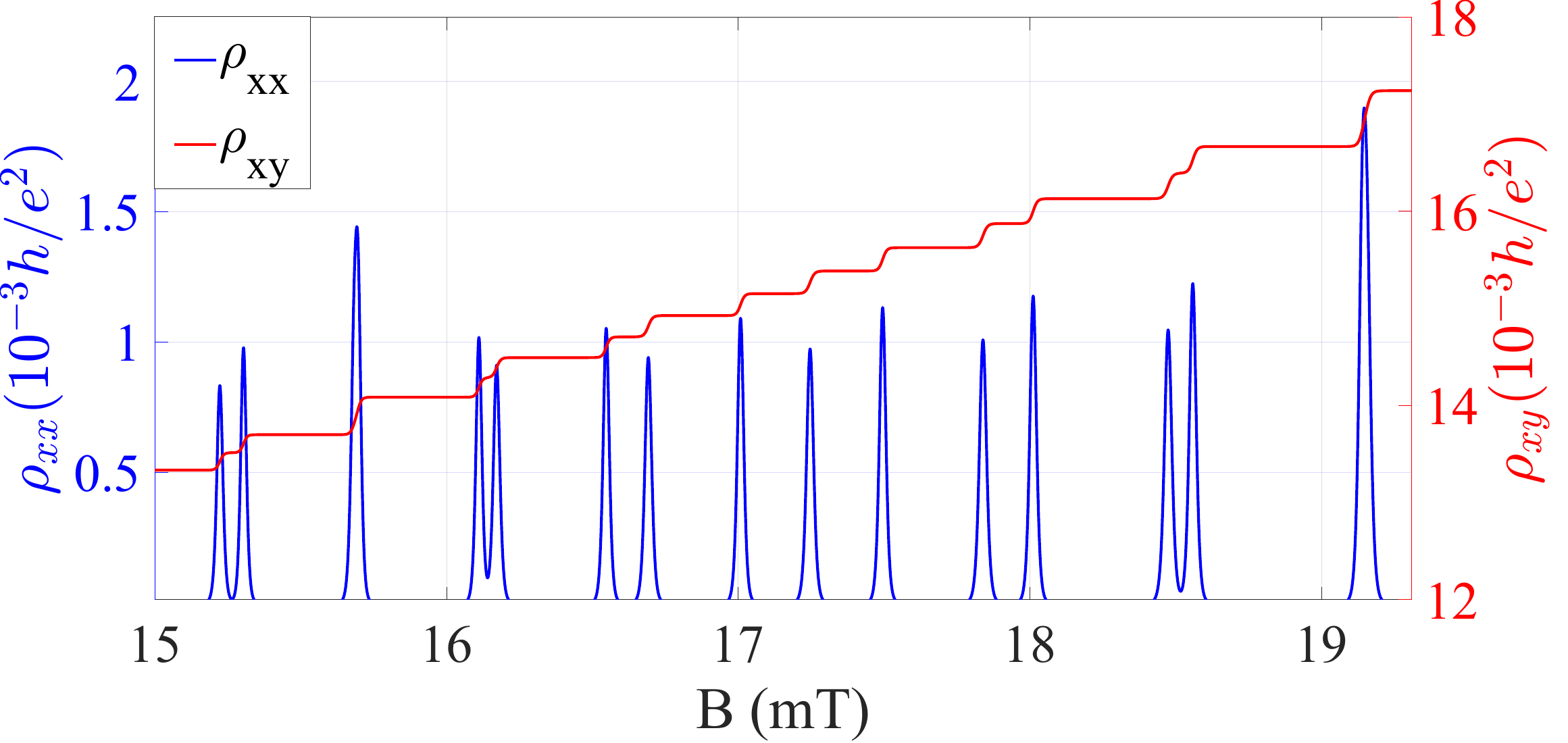}
\caption{Hall resistivity and longitudinal resistivity are plotted versus magnetic field. The Fermi level is kept at $E_F=6$ meV and $T=50$ mK. Here, the temperature is kept slightly high to restrict the height of the longitudinal peaks.}
  \label{both}
 \end{figure}

We have noted previously that intra- and/or inter-spin Landau level crossings occur with varying magnetic field. Now we would like to see how such crossing manifests itself through the Hall conductivity. The magnetic field corresponding to these crossings can be estimated by using the following condition: \(E_{n,s_1s_2}(B) = E_{n',s_1,s_2}( B)\). One such crossing is as $E_{3,\uparrow,+}(1.42 T) = E_{8,\downarrow,+}(1.42 T)$ which is reflected as a double jump in the Hall conductivity as shown in Fig.~(\ref{H1}), marked by an oval-shaped ring. This can be understood from the fact that at the same magnetic field, LLs with the same band but with the opposite spin texture acquire a step in the Hall conductivity at the same magnetic field and resulting into a double jump. Here, the Fermi level has to be adjusted very precisely at the Landau level crossing point. Any further increase in temperature will convert a straight jump into a smooth jump between two successive quantizations. 
 Now it is convenient to examine the consequences of such LL crossings on longitudinal conductivity. However, for better visualisation we plot longitudinal and Hall resistivity versus magnetic field in the same frame in Fig.~(\ref{both}) by using resistivity tensor as
 \begin{equation}
{\bf \rho}=\sigma^{-1}=\left[\begin{array}{cc}
\rho_{xx}&\rho_{xy}\\
\rho_{yx}&\rho_{yy}\end{array}\right]
 \end{equation}
 where $\rho_{xx}=\sigma_{yy}/(\sigma_{xx}\sigma_{yy}-\sigma_{xy}\sigma_{yx})$ and $\rho_{xy}=\sigma_{xy}/(\sigma_{xx}\sigma_{yy}-\sigma_{xy}\sigma_{yx})$. 
 Here, we observe the usual resistivity peaks occurs at the Hall conductivity jump. This is expected as the LLs pass through the Fermi level, resulting one jump in Hall step as well as a sharp peak in longitudinal resistivity. We note that the peak is quite larger when the Hall conductivity acquires a double jump. This corresponds to the crossing of two LLs belonging to two different bands. Hence, we can conclude that the LL crossing can also be reflected in the appearance of relatively larger height in longitudinal conductivity.
 
\begin{figure}[h!]
    \centering 
    \begin{subfigure}{0.48\textwidth}
        \centering
        \includegraphics[width=\linewidth]{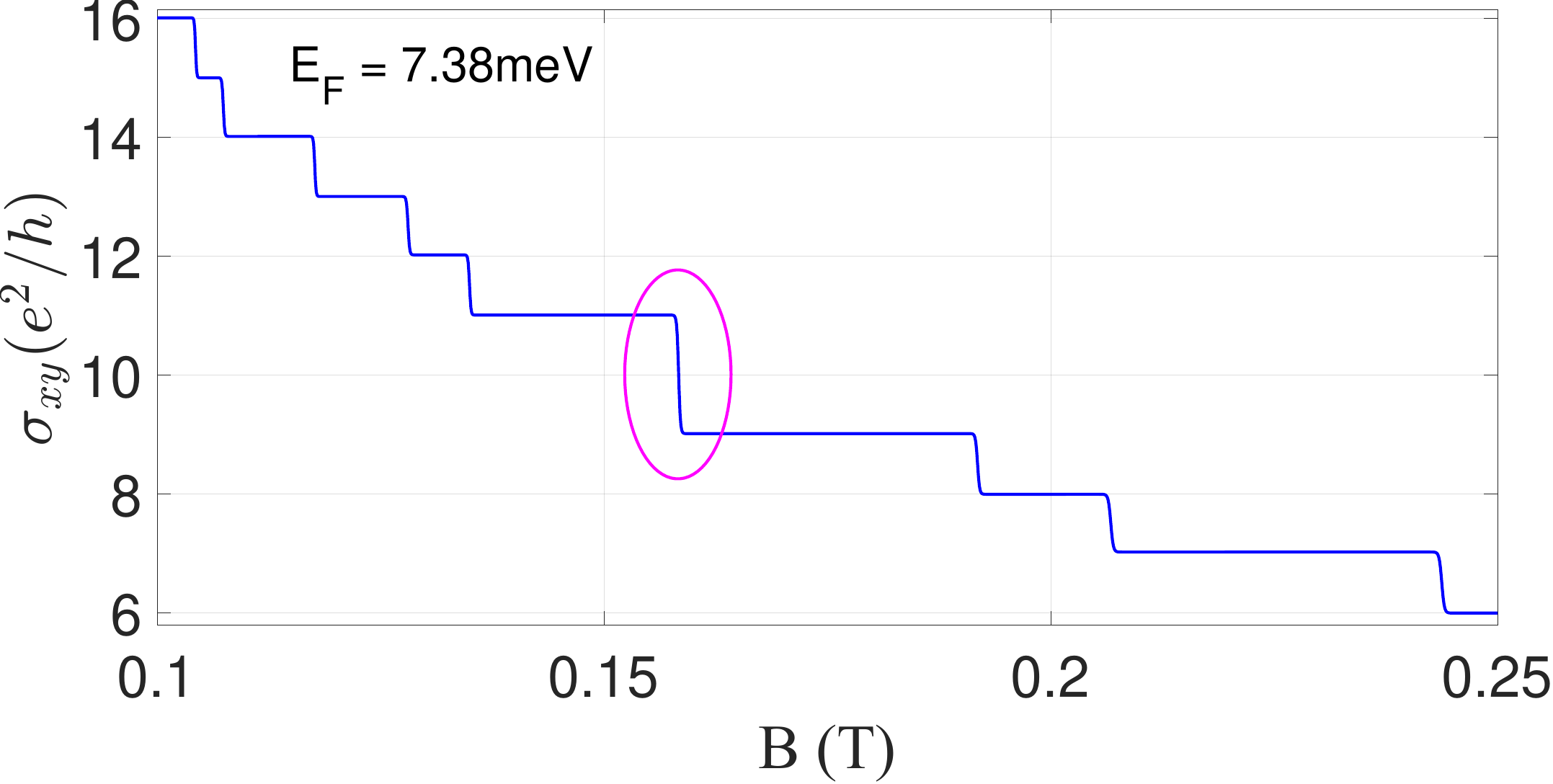}
        \caption{}
        \label{fig:hall_34}
    \end{subfigure}
    \hfill  
    \begin{subfigure}{0.48\textwidth}
        \centering
        \includegraphics[width=\linewidth]{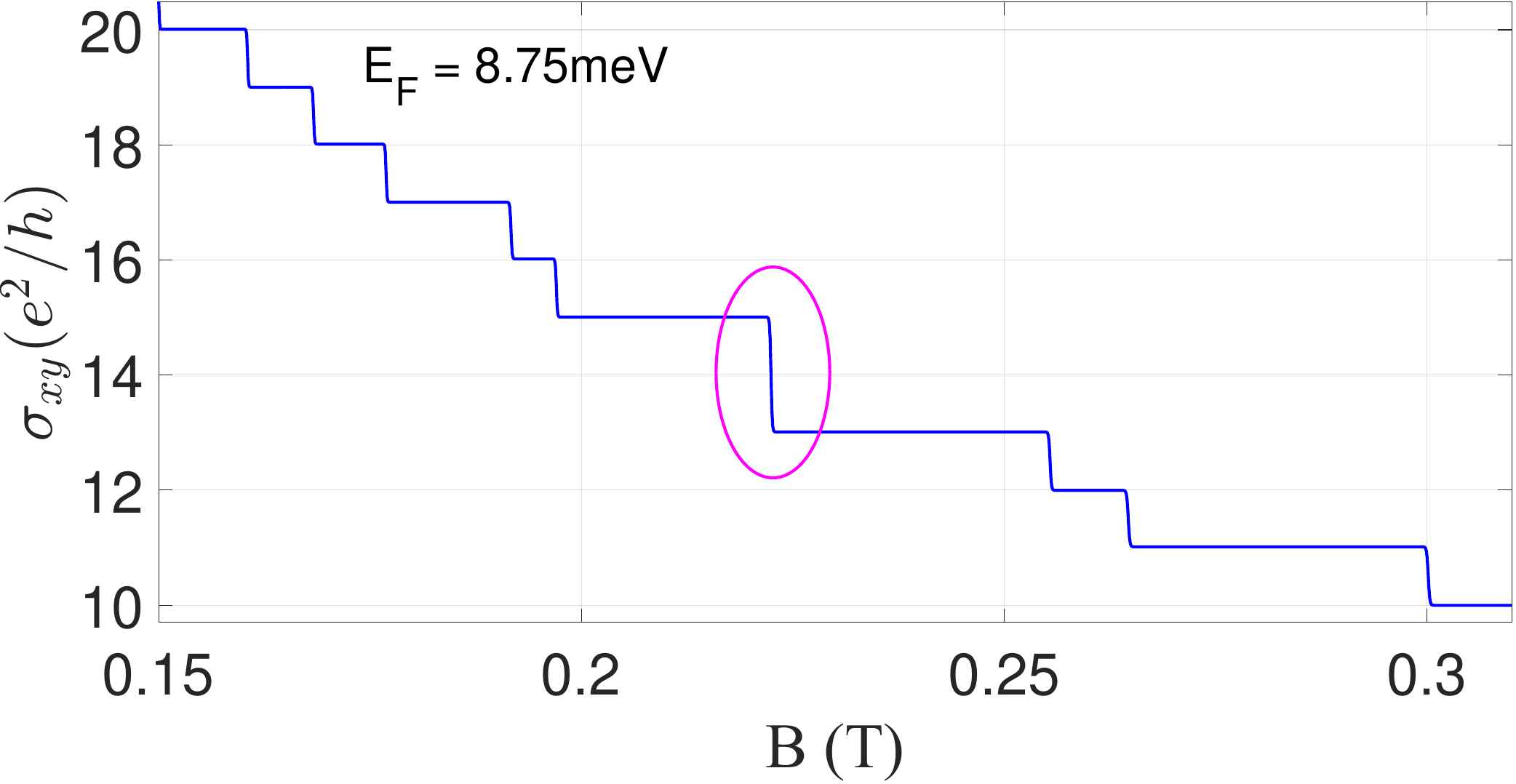}
        \caption{}
        \label{fig:hall_45}
    \end{subfigure}
        \caption{The Hall conductivity versus magnetic field for two different sets of parameter: (a) $(n,n')=(4,6)$ and $E_F=7.38$ meV (b) $(n,n')=(6,8)$ and $E_F= 8.75$ meV.}
    \label{sigma_xy1}
\end{figure}

\begin{figure}[h!]
    \centering 
    \begin{subfigure}{0.48\textwidth}
        \centering
        \includegraphics[width=\linewidth]{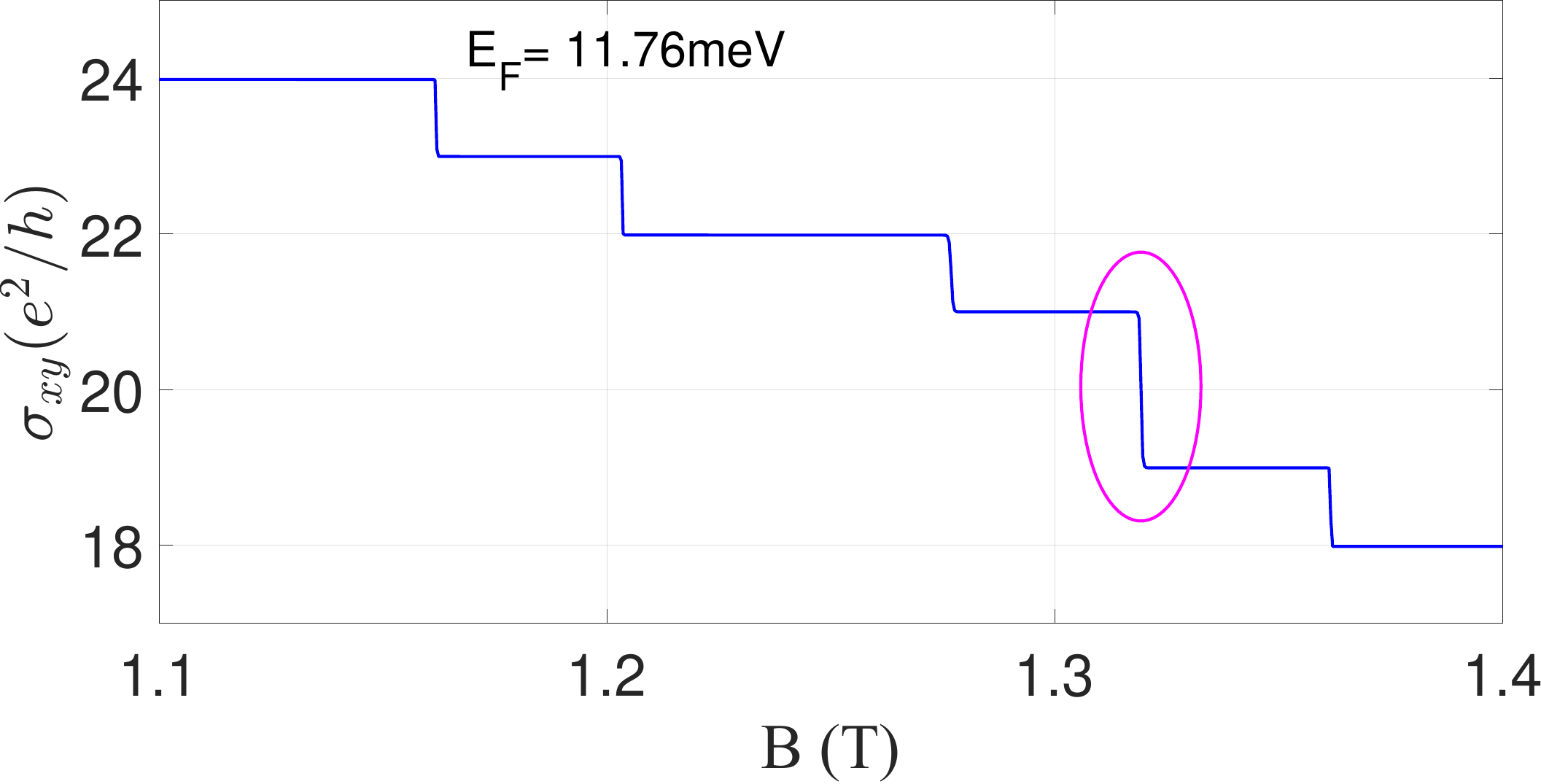}
        \caption{}
        \label{fig:hall_73}
    \end{subfigure}
    \hfill 
    \begin{subfigure}{0.48\textwidth}
        \centering
        \includegraphics[width=\linewidth]{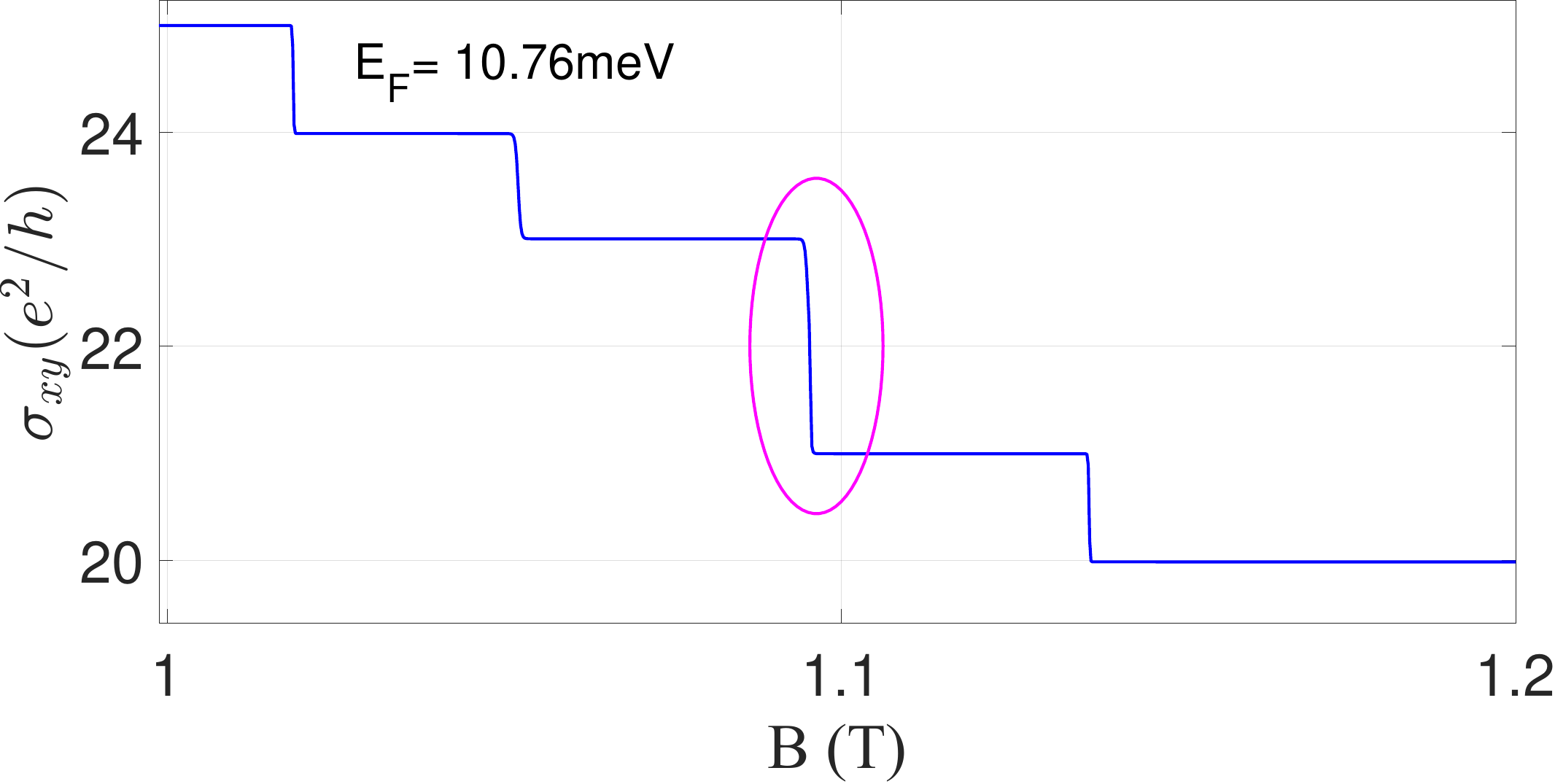}
        \caption{}
        \label{fig:hall_83}
    \end{subfigure}
    
    \caption{The Hall conductivity versus magnetic field for two different sets of parameter: (a) $(n,n')=(3,7)$ and $E_F=11.76$ meV (b) $(n,n')=(3,8)$ and $E_F= 10.76$ meV.}
    \label{sigma_xy2}
\end{figure}

 Now we give few more plots of double jump for another sets of parameter: $\alpha=5\times10^{-12}$ eV-m and $\eta=6$ meV. First we consider the inter-band LLs crossing, for which the Hall conductivity is shown in Fig.~\ref{sigma_xy1}a and Fig.~\ref{sigma_xy1}b.  In these two figures, the LL indexes ($n$,$n'$)=(4,6) and (6,8) are taken to be different. The magnetic field corresponding to these crossings are estimated by 
  \begin{align}
E_{4,\uparrow,+}(0.158) &= E_{6,\uparrow,-}(0.158) \label{hall_1} \\
E_{6,\uparrow,+}(0.222) &= E_{8,\uparrow,-}(0.222) \label{hall_2}. 
\end{align}
It can be seen from the Fig.~\ref{sigma_xy1}a and  Fig.~\ref{sigma_xy1}b that usual perfect quantization in units of $e^2/h$ occurs corresponding to each LLs, however this phenomena is abruptly interrupted by the emergence of a double jump exactly at $0.16$T and $0.22$T as estimated from the above equation, indicating the occurrence of LLs crossing. 
 Similar crossing also occurs between two opposite spin branches with the different LL indexes, and the magnetic field for inter-spin crossing can be also estimated following the same fashion as (for $(n,n')=(3,7)$ and (3,8))
 \begin{align}
E_{3,\uparrow,-}(1.318) &= E_{7,\downarrow,-}(1.318) \label{hall_3} \\
E_{3,\uparrow,-}(1.097) &= E_{8,\downarrow,-}(1.097) \label{hall_4}
\end{align}
The Hall conductivity showing the double jump corresponding to the inter-spin LLs crossing is shown in Fig.~\ref{sigma_xy2}.  The double jump appears at exactly around $B=1.31$T and $1.09$T as estimated from the above conditions.\\

Finally we comment on the possible effects of disorderedness and temperature variation \cite{PhysRevB.111.195305}. The SdH oscillation is the direct manifestation of Landau level formation. This oscillation will remain unaffected by the inclusion of disorderedness in addition to ionised impurities as long as Landau levels are well separated. The increase of disorderedness will further increase the width of the LLs, resulting the broadening of the conductivity peaks. However, a system with less disorderedness and ionised impurities is recommended to ensure the LL broadening does not destroy the separation between two successive LLs and subsequent oscillation feature.

\section{Conclusion}\label{Conclusion}
We studied the magnetotransport properties of a $2$D electronic system with unconventional RSOI. We obtained Landau levels analytically and analyzed the corresponding DOS in the presence of impurity-induced Landau level broadening. We show that LL crossing between two bands with the same spin and between two opposite spin-branches occurs with varying magnetic field. Such crossing can be attributed to the competition between band splitting and spin splitting. We show that DOS exhibits SdH oscillation with a beating pattern at the low magnetic field regime. Such beating is due to the closely separated frequencies coming from two bands in each spin branch. This Beating pattern is suppressed at high magnetic fields because of the large frequency difference between the SdH oscillations of two bands and is replaced by the usual SdH oscillations. We then employed linear reponse theory based Kubo formalism to evaluate conductivity tensor: longitudinal and quantum Hall conductivity. The longitudinal conductivity directly manifests the DOS oscillation also shows the SdH oscillation with varying magnetic field with beating pattern. We show that because of the energy separation between two spin branches, even at the $\Gamma$ point, purely spin-polarized longitudinal conductivity can be achieved by suitably placing the Fermi level.  We also plotted the Hall conductivity as a function of the Fermi level. We obtained the quantization of $e^2/h$ corresponding to each LL. However, interestingly, we note that the Hall conductivity exhibits a double jump at the LL crossing point, provided the Fermi level is also placed at that point. To examine the consequences of Landau level crossing, we estimated the magnetic field at which inter and/or intra-spin LL crossing occurred.

\section{acknowledgement}
SK Firoz Islam acknowledges the project: ANRF/ECRG/2024/005166/PMS for financial support.

\section{Data availability}
The data that support the findings of this article are not publicly available. The data are available from the authors upon reasonable request.

\begin{appendix}
\section{Wavefunction for \(n>0\)}

The coefficients \(a^{s_1,s_2}_{n}, b^{s_1,s_2}_{n}, c^{s_1,s_2}_{n}, d^{s_1,s_2}_{n}\) are given by
\begin{align}
a^{s_1,s_2}_{n} &= 
\frac{i \eta \, c^{s_1,s_2}_{n} + \beta \sqrt{n}\,\bigl(d^{s_1,s_2}_{n} + 1\bigr)}
     {\hbar\omega_c \!\left(n - \tfrac{1}{2}\right) - E_{s_1,s_2,n}}, \nonumber
b^{s_1,s_2}_{n}&= 1
\end{align}

\begin{widetext}
\begin{eqnarray}
c^{s_1,s_2}_{n} &=& 
\frac{C_0(n,s_1,s_2) D_2(n,s_1,s_2) - C_2(n,s_1,s_2) D_0(n,s_1,s_2)}
     {C_1(n,s_1,s_2) D_2(n,s_1,s_2) + C_2(n,s_1,s_2) D_1(n,s_1,s_2)}, \nonumber\\
d^{s_1,s_2}_{n} &=& 
\frac{-C_0(n,s_1,s_2) D_1(n,s_1,s_2) - C_1(n,s_1,s_2) D_0(n,s_1,s_2)}
     {C_2(n,s_1,s_2) D_1(n,s_1,s_2) + C_1(n,s_1,s_2) D_2(n,s_1,s_2)}\nonumber.  
\end{eqnarray}
\end{widetext}

The auxiliary coefficients are
\begin{eqnarray}
C_0(n,s_1,s_2) &= &\bigl[\hbar\omega_c (n-\tfrac{1}{2}) - E_{n,s_1,s_2}\bigr]\nonumber\\&\times& \bigl[\hbar\omega_c (n+\tfrac{1}{2}) - E_{n,s_1,s_2}\bigr]  - \beta^2 n, \nonumber\\
C_1(n,s_1,s_2) &= &-\beta \sqrt{n}\, \bigl[\hbar\omega_c (n-\tfrac{1}{2}) - E_{n,s_1,s_2} + i\eta\bigr],\nonumber\\
C_2(n,s_1,s_2) &=& i\eta \bigl[\hbar\omega_c (n-\tfrac{1}{2}) - E_{n,s_1,s_2}\bigr] - \beta^2 n, \nonumber\\
D_0(n,s_1,s_2) &=& \eta^2- \beta^2 n, \nonumber\\
D_1(n,s_1,s_2) &=& \beta \sqrt{n}\, \bigl[\hbar\omega_c (n-\tfrac{1}{2}) - E_{n,s_1,s_2} - i\eta\bigr], \nonumber\\
D_2(n,s_1,s_2) &=& i\eta \bigl[\hbar\omega_c (n+\tfrac{1}{2}) - E_{n,s_1,s_2}\bigr] - \beta^2 n.\nonumber
\end{eqnarray}

\section{Matrix Elements of $\mathcal{V}_x$ and $\mathcal{V}_y$}

\begin{flushleft}
For \(n > 0\), the matrix elements of $\mathcal{V}_x$ are given by:
\begin{equation}
\langle n|\mathcal{V}_x|n'\rangle 
= A_{n,n'}^{s_1,s_1',s_2,s_2'}\,\delta_{n,n'+1} + B_{n,n'}^{s_1,s_1',s_2,s_2'}\,\delta_{n,n'-1}.
\end{equation}
\begingroup\small
\begin{align*}
A_{n,n'}^{s_1,s_1',s_2,s_2'} 
&= \frac{\hbar}{\sqrt{2}ml_c} \Bigl[
    a_{n'}^{s_1',s_2'} (a_{n}^{s_1,s_2})^* \sqrt{n'-1}
  + b_{n'}^{s_1',s_2'} (b_{n}^{s_1,s_2})^* \sqrt{n'} \\[0.5ex]
&\qquad\quad
  + c_{n'}^{s_1',s_2'} (c_{n}^{s_1,s_2})^* \sqrt{n'-1}
  + d_{n'}^{s_1',s_2'} (d_{n}^{s_1,s_2})^* \sqrt{n'}
  \Bigr] \\[0.8ex]
&\quad
 - \frac{\alpha}{\hbar} \Bigl[
    a_{n'}^{s_1',s_2'} (b_{n}^{s_1,s_2})^*
  + (b_{n}^{s_1,s_2})^* c_{n'}^{s_1',s_2'}) \\[0.5ex]
&\qquad\quad
  + c_{n'}^{s_1',s_2'} (d_{n}^{s_1,s_2})^*
  + (d_{n}^{s_1,s_2})^* a_{n'}^{s_1',s_2'})
  \Bigr],
\\[2ex]
B_{n,n'}^{s_1,s_1',s_2,s_2'} 
&= -\frac{\hbar}{\sqrt{2}ml_c} \Bigl[
    (a_{n}^{s_1,s_2})^* a_{n'}^{s_1',s_2'} \sqrt{n'}
  + (b_{n}^{s_1,s_2})^* b_{n'}^{s_1',s_2'} \sqrt{n'+1} \\[0.5ex]
&\qquad\quad
  + (c_{n}^{s_1,s_2})^* c_{n'}^{s_1',s_2'} \sqrt{n'}
  + (d_{n}^{s_1,s_2})^* d_{n'}^{s_1',s_2'} \sqrt{n'+1}
  \Bigr] \\[0.8ex]
&\quad
 + \frac{\alpha}{\hbar} 
 \Bigl[
    (a_{n}^{s_1,s_2})^* b_{n'}^{s_1',s_2'}
  + b_{n'}^{s_1',s_2'} (c_{n}^{s_1,s_2})^* \\[0.5ex]
&\qquad\quad
+ (c_{n}^{s_1,s_2})^* d_{n'}^{s_1',s_2'} + d_{n}^{s_1,s_2} (a_{n}^{s_1,s_2})^*
  \Bigr].
\end{align*}
\endgroup
\end{flushleft}
\begin{adjustwidth}{-0.8cm}{-0.8cm}
\hspace{1cm}The matrix elements of $\mathcal{V}_y$ are:
\begin{equation}
\langle n|\mathcal{V}_y|n'\rangle = C_{n,n'}^{s_1,s_1',s_2,s_2'}\,\delta_{n,n'+1} + D_{n,n'}^{s_1,s_1',s_2,s_2'}\,\delta_{n,n'-1}.\nonumber
\end{equation}

\vspace{0.3cm}
\begingroup\small
\begin{align*}
C_{n,n'}^{s_1,s_1',s_2,s_2'} 
&= \frac{\hbar}{\sqrt{2}ml_c} \Bigl[
   (a^{s_1',s_2'}_{n'})^* a^{s_1,s_2}_{n} \sqrt{n}
 + b^{s_1,s_2}_{n} (b^{s_1',s_2'}_{n'})^* \sqrt{n+1} \\[0.5ex]
&\qquad\quad
 + c^{s_1,s_2}_{n} (c^{s_1',s_2'}_{n'})^* \sqrt{n}
 + (d^{s_1',s_2'}_{n'})^* d^{s_1,s_2}_{n} \sqrt{n+1}
 \Bigr] \\[0.8ex]
&\quad
 - \frac{\alpha}{\hbar} \Bigl[
   (a^{s_1',s_2'}_{n'})^* b^{s_1,s_2}_{n}
 + b^{s_1,s_2}_{n} (c^{s_1',s_2'}_{n'})^* \\[0.5ex]
&\qquad\quad
 + (c^{s_1',s_2'}_{n'})^* d^{s_1,s_2}_{n}
 + d^{s_1,s_2}_{n} (a^{s_1',s_2'}_{n'})^*
 \Bigr],
\\[3ex]
\hspace{0.5cm}
D_{n,n'}^{s_1,s_1',s_2,s_2'} 
&= -\frac{\hbar}{\sqrt{2}ml_c} \Bigl[
   a^{s_1,s_2}_{n} (a^{s_1',s_2'}_{n'})^* \sqrt{n-1}
 + b^{s_1,s_2}_{n} (b^{s_1',s_2'}_{n'}))^* \sqrt{n} \\[0.5ex]
&\qquad\quad
 + c^{s_1',s_2'}_{n'} (c^{s_1,s_2}_{n})^* \sqrt{n-1}
 + d^{s_1,s_2}_{n} (d^{s_1',s_2'}_{n'})^* \sqrt{n}
 \Bigr] \\[0.8ex]
&\quad
 - \frac{\alpha}{\hbar} \Bigl[
   a^{s_1,s_2}_{n} (b^{s_1',s_2'}_{n'})^*
 + (b^{s_1',s_2'}_{n'})^* c^{s_1,s_2}_{n} \\[0.5ex]
&\qquad\quad
 + c^{s_1,s_2}_{n} (d^{s_1',s_2'}_{n'})^*
 + (d^{s_1',s_2'}_{n'})^* a^{s_1,s_2}_{n}
 \Bigr].
\end{align*}
\endgroup
\end{adjustwidth}

For \(n=0\), the matrix product becomes:
\begin{align}
Q_{10} &= \langle 1 | \mathcal{V}_x | 0 \rangle 
          \, \langle 0 | \mathcal{V}_y | 1 \rangle.\nonumber
\end{align}

\begingroup
\small
\begin{align*}
\langle 1 | \mathcal{V}_x | 0 \rangle \nonumber
&= \frac{\alpha}{\hbar} \Bigl[ 
    (a^{s_1,s_2}_{1})^* b^{s_1,s_2}_{0}
  + (a^{s_1,s_2}_{1})^* d^{s_1,s_2}_{0} \\[0.5ex]
&\qquad\quad
  + b^{s_1,s_2}_{0} (c^{s_1,s_2}_{1})^*
  + (c^{s_1,s_2}_{1})^* d^{s_1,s_2}_{0}
  \Bigr] \\[0.5ex]
&\quad 
 - \frac{\hbar}{\sqrt{2} m l_c} \Bigl[
   (b^{s_1,s_2}_{1})^* b^{s_1,s_2}_{0}
 + (d^{s_1,s_2}_{1})^* d^{s_1,s_2}_{0}
 \Bigr], \\[2ex]
\langle 0 | \mathcal{V}_y | 1 \rangle 
&= \frac{\alpha}{\hbar} \Bigl[
   - a^{s_1,s_2}_{1} (b^{s_1,s_2}_{0})^*
   + a^{s_1,s_2}_{1} (d^{s_1,s_2}_{0})^* \\[0.5ex]
&\qquad\quad
   - (b^{s_1,s_2}_{0})^* c^{s_1,s_2}_{1}
   - c^{s_1,s_2}_{1} (d^{s_1,s_2}_{0})^*
   \Bigr] \\[0.5ex]
&\quad 
 + \frac{\hbar}{\sqrt{2} m l_c} \Bigl[
   b^{s_1,s_2}_{1}(b^{s_1,s_2}_{0})^*
 + d^{s_1,s_2}_{1} (d^{s_1,s_2}_{0})^*
 \Bigr].
\end{align*}
\endgroup
Similarly, we obtain,
\begin{align}
Q_{01} &= \langle 0 | \mathcal{V}_x | 1 \rangle 
          \, \langle 1 | \mathcal{V}_y | 0 \rangle.\nonumber
\end{align}

\begingroup
\small
\begin{align*}
\langle 0 | \mathcal{V}_x | 1 \rangle 
&= \frac{\alpha}{\hbar} \Bigl[
   - a^{s_1,s_2}_{1} (b^{s_1,s_2}_{0})^*
   - a^{s_1,s_2}_{1} (d^{s_1,s_2}_{0})^* \\[0.5ex]
&\qquad\quad
   - (b^{s_1,s_2}_{0})^* c^{s_1,s_2}_{1}
   - c^{s_1,s_2}_{1} (d^{s_1,s_2}_{0})^*
   \Bigr] \\[0.5ex]
&\quad 
 + \frac{\hbar}{\sqrt{2} m l_c} \Bigl[
   b^{s_1,s_2}_{1} (b^{s_1,s_2}_{0})^*
 + d^{s_1,s_2}_{1} (d^{s_1,s_2}_{0})^*
 \Bigr], \\[2ex]
\langle 1 | \mathcal{V}_y | 0 \rangle 
&= \frac{\alpha}{\hbar} \Bigl[
   - (a^{s_1,s_2}_{1})^* b^{s_1,s_2}_{0}
   - (a^{s_1,s_2}_{1})^* d^{s_1,s_2}_{0} \\[0.5ex]
&\qquad\quad
   - b^{s_1,s_2}_{0} (c^{s_1,s_2}_{1})^*
   - (c^{s_1,s_2}_{1})^* d^{s_1,s_2}_{0}
   \Bigr] \\[0.5ex]
&\quad 
 - \frac{\hbar}{\sqrt{2} m l_c} \Bigl[
   (b^{s_1,s_2}_{1})^* b^{s_1,s_2}_{0}
 - (d^{s_1,s_2}_{1})^* d^{s_1,s_2}_{0}
 \Bigr].
\end{align*}
\endgroup

\end{appendix}
\bibliography{URS}

\end{document}